\documentclass[superscriptaddress,nofootinbib,eqsecnum,onecolumn]{revtex4-1}

\pdfoutput=1

\usepackage{graphicx}
\usepackage{bm}
\usepackage{dcolumn}
\usepackage{amssymb}
\usepackage{amsmath}
\usepackage{epsfig} 
\usepackage{subfigure}
\usepackage[english]{babel}
\usepackage[utf8]{inputenc}
\usepackage{eucal}
\usepackage{changepage}
\usepackage{hyperref}
\usepackage{verbatim}
\usepackage{latexsym}
\usepackage{color,xcolor}
\usepackage{ulem}
\usepackage{soul}  
\newcommand{\termchancery}{\fontfamily{pzc}\selectfont}

\hypersetup{colorlinks=true,linkcolor=blue,filecolor=magenta,urlcolor=cyan,citecolor=orange}

\begin{document}

\title{Dense and Cold Magnetized Quark Matter:
A Review of Magnetic-Field-Independent Regularization and the Medium Separation Scheme}

\author{Francisco X. Azeredo}  
\email[Electronic address: ]{francisco.azeredo@acad.ufsm.br}
\affiliation{Departamento de F\'{\i}sica, Universidade Federal de Santa Maria, 97105-900, Santa Maria, Rio Grande do Sul, Brazil}

\author{Dyana C. Duarte} 
\email[Electronic address: ]{dyana.duarte@ufsm.br }
\affiliation{Departamento de F\'{\i}sica, Universidade Federal de Santa Maria, 97105-900, Santa Maria, Rio Grande do Sul, Brazil}

\author{Ricardo L. S. Farias}  
\email[Electronic address: ]{ricardo.farias@ufsm.br}
\affiliation{Departamento de F\'{\i}sica, Universidade Federal de Santa Maria, 97105-900, Santa Maria, Rio Grande do Sul, Brazil}
\affiliation{Center for Nuclear Research, Department of Physics, Kent State University, Kent, Ohio 44242 USA} 

\author{ Bruno S. Lopes} 
\email[Electronic address: ]{bruno.lopes@acad.ufsm.br }
\affiliation{Departamento de F\'{\i}sica, Universidade Federal de Santa Maria, 97105-900, Santa Maria, Rio Grande do Sul, Brazil}
\affiliation{Center for Nuclear Research, Department of Physics, Kent State University, Kent, Ohio 44242 USA}   

\author{ Jo\~ao A. R. S. Prado} 
\email[Electronic address: ]{joao.prado@acad.ufsm.br }
\affiliation{Departamento de F\'{\i}sica, Universidade Federal de Santa Maria, 97105-900, Santa Maria, Rio Grande do Sul, Brazil}

 \author{ William R. Tavares} 
 \email[Electronic address: ]{tavares.william@ce.uerj.br }
\affiliation{Departamento de F\'{\i}sica Te\'orica, Universidade do
  Estado do Rio de Janeiro, 20550-013 Rio de Janeiro, RJ, Brazil.} 
  \affiliation{ CFisUC, Department of Physics, University of Coimbra, P-3004 - 516 Coimbra, Portugal}

\begin{abstract}

We present a comprehensive review of regularization schemes for magnetized dense quark matter within effective models of quantum chromodynamics, focusing on the Magnetic-Field-Independent Regularization (MFIR) and the Medium Separation Scheme (MSS) at finite chemical potential and magnetic field. In nonrenormalizable frameworks such as the Nambu–Jona-Lasinio model, the treatment of ultraviolet divergences is crucial, particularly in magnetized and dense environments where conventional regularization procedures may introduce unphysical artifacts. We show that MFIR consistently isolates divergent vacuum contributions from finite magnetic-field-dependent terms, while MSS extends this separation to the medium sector, ensuring that only vacuum quantities are regularized. Within this unified framework, we analyze the thermodynamics of cold and dense quark matter, including color-superconducting phases, and demonstrate that the superconducting gap remains finite at large chemical potentials, even in the presence of strong magnetic fields. In contrast to results obtained with traditional regularization schemes, we find no evidence for a transition to a normal phase at zero temperature, highlighting the importance of a proper separation between vacuum and medium contributions. These results eliminate spurious oscillations and other nonphysical artifacts, leading to a more robust and physically consistent description of strongly interacting matter under extreme conditions relevant to compact stars and heavy-ion collisions.

\end{abstract}

\maketitle


\section{Introduction}
\label{sec:Introduction}

One of the most ambitious phenomena to be confirmed by noncentral heavy-ion collision experiments is the presence of strong electromagnetic fields \cite{Rafelski:1975rf, Kharzeev:2007jp, Skokov:2009qp, Bzdak:2011yy, Ou:2011fm, Voronyuk:2011jd, Bloczynski:2012en, Deng:2012pc, Bloczynski:2013mca, Zhong:2014cda, Zhong:2014sua, Fukushima:2018grm, Hattori:2023egw, Adhikari:2024bfa, Endrodi:2024cqn, Mustafa:2025uad}. In collisions at RHIC and ALICE, these fields are expected to reach magnitudes of the order of \mbox{$eB\sim10^{18}$ G} and $eB\sim10^{19}$ G, respectively \cite{Kharzeev:2007jp,Skokov:2009qp}, thereby providing an ideal environment to investigate their effects on the properties of strongly interacting matter.
The dominant fields, as inferred from event-averaged magnetic-field estimates \cite{Deng:2012pc,Huang:2015oca}, are expected to be aligned perpendicular to the reaction plane and are generally believed to persist only for a very short time. However, owing to the effects of finite electric conductivity \cite{Arnold:2003zc, Aarts:2007wj, Ding:2010ga, Francis:2011bt, Brandt:2012jc, Amato:2013naa, Aarts:2014nba, Ding:2014dua, Huang:2015oca}, their lifetime may be significantly extended. The chiral magnetic effect \cite{Fukushima:2008xe} is the primary phenomenon to be experimentally explored, and its recent \cite{STAR:2021mii, ALICE:2022ljz, STAR:2025vhs, ALICE:2026cvp} and future achievements could lead to a better understanding of  strongly interacting matter under external fields.
While experimental programs continue to advance, it is very common to find in the literature different predictions from lattice QCD (LQCD) \cite{Bali:2011qj, Bali:2012zg, Hidaka:2012mz, Luschevskaya:2012xd, Bali:2013esa, Bali:2013owa, Endrodi:2013cs, Andreichikov:2016ayj, Bali:2017ian, DElia:2018xwo, Braguta:2019yci, Endrodi:2019zrl, Ding:2020inp, Guenther:2020jwe, DElia:2021yvk, Ding:2022tqn}, effective theories \cite{Andersen:2012zc, Andersen:2014xxa, Adhikari:2015wva, Hofmann:2017noo, Hofmann:2020dvz, Hofmann:2020lfp, Adhikari:2023fdl}, and models \cite{Klevansky:1992qe, Gatto:2010qs, Duarte:2011ph, Skokov:2011ib, Andersen:2012jf, Avancini:2012ee, Andersen:2014xxa, Ayala:2014mla, Ayala:2015lta, Magdy:2015eda, Farias:2016gmy, Braghin:2017zas, Endrodi:2019whh, Fang:2019xbk, Andersen:2021lnk, Krein:2021sco, Ali:2024mnn} that include magnetic fields in the study of phase diagrams~\cite{Gatto:2010qs, Skokov:2011ib, Andersen:2012jf, Avancini:2012ee, Andersen:2014xxa, Ferreira:2014exa, Ferreira:2014kpa, Magdy:2015eda, DElia:2018xwo, Bandyopadhyay:2019pml, Braguta:2019yci, Fang:2019xbk, Ding:2020inp, Guenther:2020jwe, Andersen:2021lnk, DElia:2021yvk, Ali:2024mnn}, meson masses \cite{Fayazbakhsh:2012vr, Hidaka:2012mz, Luschevskaya:2012xd, Ayala:2015lta, Hattori:2015aki, Andreichikov:2016ayj, Avancini:2016fgq, Aguirre:2017dht, Bali:2017ian, Mao:2017wmq, Wang:2017vtn, Aguirre:2018fbo, Avancini:2018svs, Ayala:2018zat, Coppola:2018vkw, Ayala:2020dxs, Coppola:2020mon, Ding:2020hxw, GomezDumm:2020bxj, Sheng:2020hge, Aguirre:2021ljk, Avancini:2022qcp, Carlomagno:2022inu, Das:2022mic, Ding:2022tqn, Mei:2022dkd, Coppola:2024uvz}, and thermodynamic quantities \cite{Bali:2013owa, Duarte:2015ppa, Magdy:2015eda, Farias:2016gmy, Avancini:2017gck, Aguirre:2019ivr, Aguirre:2020tiy, Avancini:2020xqe, Tavares:2021fik}.
The interest in strong magnetic fields is not limited to heavy-ion collisions; it also extends to high-density systems, where the most prominent physical objects are magnetars \cite{Duncan:1992hi,Kouveliotou:1998ze,Sinha:2026gkk}, whose magnetic fields can reach values of the order of $eB\sim10^{16}$ G. In the regime of very high densities and low temperatures, one can also investigate several applications related to the QCD phase diagram, such as the effects of magnetic fields on superconducting phases \cite{Allen:2015paa,Duarte:2015ppa}. Likewise, strong magnetic fields are also expected to play a role in the electroweak phase transition \cite{Grasso:2000wj}, which is relevant for understanding the evolution of the early Universe.

Quantum chromodynamics (QCD) is the fundamental theory of strong interactions. However, in the low-energy regime, its nonperturbative nature requires the use of nonperturbative methods.
One possible approach is to employ effective models that are simpler than QCD while preserving its most important symmetry properties, particularly chiral symmetry. These models provide a useful framework to investigate how chiral symmetry is modified under extreme conditions, such as finite temperature, baryon density, and strong magnetic fields. Among the various effective approaches available in the literature, two of the most widely employed are the Nambu--Jona-Lasinio (NJL) model \cite{Nambu:1961tp,Nambu:1961fr} and the linear sigma model with quarks \cite{Scavenius:2000qd,Andersen:2014xxa}, the latter being a quark-level generalization of the original linear sigma model \cite{Gell-Mann:1960mvl}. These models are often studied within the mean-field approximation, which considerably simplifies the investigation of partial chiral symmetry restoration through the behavior of the chiral condensate.
In the present review, we focus on the application of constant and strong magnetic fields within the framework of the NJL model at finite density.

As discussed above, symmetry considerations provide valuable guidance for understanding the various phases of QCD. In the context of chiral symmetry breaking, the chiral condensate is commonly regarded as an approximate order parameter, and its behavior under extreme conditions, such as finite temperature, density, and strong electromagnetic fields, has been extensively investigated over the last decades \cite{Klevansky:1992qe, Buballa:2003qv, Menezes:2008qt, Andersen:2014xxa}.
In the presence of magnetic fields, it is well established that the chiral condensate is enhanced, strengthening the chirally broken phase through a mechanism commonly referred to as magnetic catalysis \cite{Gusynin:1995nb,Semenoff:1999xv,Shovkovy:2012zn}. This phenomenon has been extensively investigated in lattice QCD \cite{Bali:2012zg}, effective models \cite{Menezes:2008qt}, and low-dimensional systems relevant to condensed matter physics \cite{Ramos:2013aia}. However, in the vicinity of the pseudocritical temperature, LQCD predicts the inverse magnetic catalysis effect \cite{Bali:2012zg,Bandyopadhyay:2020zte}, characterized by a nonmonotonic dependence of the chiral condensate on the magnetic field strength. Such behavior is not naturally reproduced by conventional effective models, thereby introducing an important constraint that must be properly accounted for in low-energy descriptions of QCD. Several approaches have been proposed to address this puzzle, including the introduction of magnetic-field-dependent couplings, which can be determined through different fitting procedures available in the literature \cite{Farias:2014eca, Ferreira:2014kpa, Avancini:2016fgq, Endrodi:2019whh, Moreira:2020wau}; the implementation of a magnetic-field-dependent $T_0$ parameter in Polyakov-loop potentials, where $T_0$ represents the pseudocritical temperature associated with the deconfinement transition \cite{Ferreira:2013tba}; or approaches based on analytical developments \cite{Ayala:2020muk}, including beyond-mean-field treatments~\cite{Fukushima:2012kc,Mao:2016fha} and the incorporation of the anomalous magnetic moment of quarks \cite{Fayazbakhsh:2014mca}.

The Nambu--Jona-Lasinio model is a particularly useful framework for investigating symmetry aspects and phase transitions, as it is considerably simpler than QCD while remaining applicable to a wide variety of physical environments. However, the model is nonrenormalizable, and ultraviolet divergent integrals must be regularized in order to obtain finite physical results.  Several regularization schemes have been explored in the literature \cite{Avancini:2018svs}, including the 3D sharp cutoff, 4D sharp cutoff, Proper-Time, and Pauli--Villars regularizations, among others. The regularization procedure introduces an ultraviolet cutoff, $\Lambda$, which becomes one of the fundamental parameters of the NJL model, together with the coupling constant, $G$, and the current quark mass, $m_0$. These parameters are fixed by reproducing the vacuum values of the pion mass, the pion decay constant, and the chiral condensate. Despite its simplicity, the lack of renormalizability can strongly influence the resulting physical predictions \cite{Avancini:2019wed,Avancini:2020xqe}, particularly when the model is studied in the presence of constant external electromagnetic fields.
For the purposes of the present review, it is useful to classify the regularization procedures into two broad categories, according to how they treat the divergent contributions associated with the magnetic field:

\begin{itemize}

\item
MFIR (Magnetic-Field-Independent Regularization): A class of regularization procedures that completely separates vacuum contributions from magnetic-field-dependent contributions \cite{Ebert:1999ht, Ebert:2003yk, Allen:2015paa, Duarte:2015ppa, Avancini:2019wed, Avancini:2020xqe}. Within this framework, only the vacuum contribution is regularized.

\item nonMFIR: Regularization schemes that do not separate vacuum and magnetic-field-dependent contributions \cite{Avancini:2020xqe}. In this case, the complete expression, containing both external-field effects and vacuum contributions, is regularized.

\end{itemize}

The central idea of the MFIR procedure is to construct, for instance, an effective potential in which the vacuum contribution, which originally depends on the ultraviolet cutoff of the theory, is explicitly separated from the magnetic-field-dependent contributions, which are finite by construction.
This procedure is based on the separation of divergences originally introduced by Schwinger \cite{Schwinger:1951nm} in the context of quantum electrodynamics and has been extensively adopted in QCD-inspired models over the last decades.
Nevertheless, there exists a class of regularization prescriptions that does not separate vacuum contributions from magnetic-field-dependent terms, which we collectively denote as nonMFIR. Within such schemes, one typically encounters divergent contributions to the effective potential or gap equations that simultaneously depend on both the magnetic field and the ultraviolet cutoff.
This entanglement between vacuum and magnetic-field contributions in divergent integrals can generate several unphysical effects, including oscillations in the chiral condensate as a function of the magnetic field \cite{Avancini:2019wed}, tachyonic meson masses~\cite{Fayazbakhsh:2012vr}, artificial de Haas--van Alphen oscillations in the magnetization of high-density systems~\cite{Allen:2015paa,Duarte:2015ppa,Sinha:2015bva}, and even qualitatively different phase-transition patterns when the quark anomalous magnetic moment is taken into account \cite{Farias:2021fci,Tavares:2023oln}. In this sense, MFIR-based schemes are particularly attractive because they eliminate the unphysical artifacts discussed above and, in ge\-ne\-ral, lead to results that are in better agreement with lattice QCD calculations. An extension of the MFIR framework, known as vacuum magnetic regularization~\cite{Avancini:2020xqe,He:2024gnh}, has proven particularly useful for the calculation of quantities directly comparable to lattice QCD results, such as the renormalized magnetization \cite{Avancini:2020xqe}. Consequently, field-independent regularization procedures have become valuable tools for investigating important physical phenomena even beyond the scope of current lattice simulations, including high-density systems \cite{Duarte:2015ppa,Allen:2015paa,He:2022inw}. Analogous procedures that avoid mixing vacuum and medium contributions have also demonstrated the ability to produce results that are more closely aligned with lattice QCD predictions. A notable example is the medium separation scheme (MSS), a regularization framework inspired by the findings of Ref.~\cite{Battistel:1998tj}, which has been successfully applied to the study of chiral imbalance \cite{Farias:2016let}, isospin QCD \cite{Lopes:2025rvn}, two-color dense QCD \cite{Pasqualotto:2025kpo}, and color superconductivity \cite{Farias:2005cr,Azeredo:2026wgl}. Motivated by these developments, we revisit the fundamental ideas underlying MFIR-type procedures in the context of the NJL model with color superconductivity.

Asymptotic freedom implies that quark matter becomes weakly interacting at energy scales sufficiently larger than the characteristic QCD scale. Consequently, at zero temperature and sufficiently high baryon densities, a close analogy with the Bardeen--Cooper--Schrieffer (BCS) theory of superconductivity emerges: quark excitations near the Fermi surface dominate the dynamics, and attractive interactions render the formation of paired states energetically favorable. The resolution of this instability occurs through Cooper pairing, which, in the context of QCD, corresponds to the formation of a diquark condensate. Such a condensate breaks color symmetry and defines the true asymptotic ground state of dense quark matter. This phenomenon, known as color superconductivity, was proposed in the early stages of the development of QCD~\cite{Collins:1974ky,Barrois:1977xd,Frautschi:1978rz,Bailin:1983bm} and has remained the subject of extensive theoretical investigation ever since~\cite{Alford:1997zt, Alford:1998mk, Berges:1998rc, Alford:2001dt, Buballa:2003qv, Huang:2003xd, Shovkovy:2003uu, Alford:2007xm}. Interest in this subject increased substantially after pioneering estimates of the superconducting gap suggested values considerably larger ($\sim 100$~MeV) than originally anticipated~\cite{Alford:1997zt,Rapp:1997zu}, indicating that color-superconducting phases may survive up to relatively high temperatures. A comprehensive mapping of these phases within the QCD phase diagram has been developed over the years~\cite{Alford:2007xm}, ranging from asymptotically dense regimes accessible through first-principles calculations to moderate-density regions that require phenomenological modeling. At asymptotically high densities, quark matter is expected to reside in the Color--Flavor--Locked (CFL) phase, where color and chiral symmetries are spontaneously broken down to a residual diagonal SU(3) subgroup~\cite{Alford:1998mk}. However, because the strange quark mass remains significantly larger than those of the light flavors, an intermediate-density window is expected to exist in which only up and down quarks participate in Cooper pairing, resulting in only a partial breaking of color symmetry. This regime is known as the two-flavor color-superconducting (2SC) phase and constitutes the primary focus of the present work. At intermediate baryon densities, such as those expected in the cores of compact stars, where matter may reach densities as high as ten times the nuclear saturation density $\rho_0 \approx 0.16\text{ fm}^{-3}$, the system is generally expected to be described by the two-flavor color-superconducting phase (2SC). For comprehensive reviews of this regime and discussions of asymptotic first-principles calculations, we refer the reader to Refs.~\cite{Rajagopal:2000wf,Alford:2001dt,Rischke:2003mt} and~\cite{Son:1998uk,Shovkovy:1999mr}, respectively.

Numerous studies demonstrate how thermal and density scales shape a rich and intricate QCD phase diagram~\cite{Alford:1998mk,Schmitt:2003xq,Duarte:2018kfd}.
Furthermore, the transport properties of the 2SC phase have direct astrophysical implications. For example, a recent self-consistent NJL-model study demonstrated that electron and muon neutrino absorption in equilibrated 2SC matter may generate a degenerate neutrino gas with a mean free path of only a few meters, a feature of potential relevance for neutron-star merger dynamics~\cite{Alford:2025jtm}.
In this context, the construction of thermal equations of state suitable for numerical simulations of neutron-star mergers requires efficient and reliable theoretical frameworks. Recently, a finite-temperature formalism was developed to systematically extend cold quark-matter equations of state to finite temperatures while consistently incorporating the suppression of thermal pressure in color-superconducting phases such as 2SC and CFL~\cite{Gholami:2025yqf}. When magnetized compact objects such as magnetars are considered, the presence of intense background magnetic fields introduces additional layers of complexity into the description of color-superconducting quark matter. Under such extreme conditions, certain non-spherical superconducting phases may not exhibit the conventional electromagnetic Meissner effect, allowing magnetic flux to penetrate the medium without an associated energetic cost~\cite{Feng:2009vt}. Consequently, strong magnetic fields can substantially modify both the pairing dynamics and the resulting equation of state~\cite{Abhishek:2018xml}. In the two-flavor sector, this interplay requires the implementation of electric and color neutrality constraints in order to properly account for rotated charges and anisotropic superconducting properties~\cite{Cao:2015xja,Coppola:2017edn}. Although strong magnetic fields are well known to induce de Haas--van Alphen-like oscillations in the superconducting gap parameters~\cite{Fukushima:2007fc,Noronha:2007wg}, they can also trigger more exotic phenomena. Examples include the formation of inhomogeneous gluon condensates capable of removing chromomagnetic instabilities in neutral 2SC matter, as well as the generation of rotated magnetic fields~\cite{Ferrer:2006ie,Ferrer:2007uw,Yuan:2024ajk}. In three-flavor matter, strong magnetic fields can further modify the collective excitation spectrum and favor the emergence of novel phases, such as the magnetic color--flavor--locked and paramagnetic color--flavor--locked phases, in which aligned Cooper-pair magnetic moments enhance the pairing energy~\cite{Ferrer:2007iw,Paulucci:2010uj,Feng:2011fj}. From a more formal perspective, renormalization-group analyses at asymptotically high densities reveal that long-range magnetic interactions lead to a non-BCS scaling behavior of the superconducting gap, namely $\Delta \sim \mu g^{-5} \exp(-c/g)$~\cite{Son:1998uk,Hsu:1999mp}. Finally, although it has been conjectured that sufficiently strong magnetic fields could induce vacuum electromagnetic superconductivity through the condensation of charged vector $\rho$ mesons~\cite{Chernodub:2010qx,Chernodub:2012tf}, recent calculations based on extended NJL models that consistently incorporate Schwinger phases and $B \neq 0$ wave functions have found no evidence supporting this mechanism, thereby disfavoring its realization under physically relevant conditions~\cite{Carlomagno:2022inu,Carlomagno:2022arc}.

The emergence of a nonvanishing superconducting gap has important consequences for the equation of state of dense matter~\cite{Rajagopal:2000wf,Alford:2001dt}.
In this regard, Ref.~\cite{Kurkela:2024xfh} argues that current astrophysical constraints derived from observations of pulsars and neutron-star mergers impose upper bounds on the color-superconducting gap in the CFL phase, estimated to be $\Delta \sim 457$~MeV or $\Delta \sim 216$~MeV, depending on the strictness of the assumptions employed. \textcolor{black}{Recent works calculated next-to-leading-order corrections to CFL pressure under neutron-star equillibrium conditions to constrain the diquark gap to $\Delta \lesssim 140$ MeV~\cite{Geissel:2025vnp}, whereas, combining multi-messenger observations with perturbative QCD and chiral effective field theory in Bayesian inference framework, it was obtained a 95\% credible upper limit about 66~MeV~\cite{Tang:2026yta}.}  Effective models have also been extensively explored in this context. In particular, massive hybrid stars can be successfully described both within the quark--meson--diquark model~\cite{Andersen:2026sjc}, and through a relativistic density-functional approach that mimics confinement via medium-dependent couplings and incorporates diquark pairing~\cite{Ivanytskyi:2022oxv}, as well as within the NJL model supplemented by renormalization group consistency requirements~\cite{Gholami:2024ety}, with vector interactions included in both frameworks. In both approaches, the high-density sector is treated with particular care, leading to a nonvanishing color-superconducting phase throughout the relevant density range~\cite{Gholami:2024diy,Andersen:2026xrf}. In the present review, we argue that this scenario provides the most physically consistent description of dense quark matter and that the disappearance of the superconducting gap observed in some studies may be associated with cutoff artifacts, which can be systematically controlled through the medium separation scheme. Additional studies of compact stars based on effective models can be found in Refs.~\cite{Grigorian:2003vi,Blaschke:2008vh,Bonanno:2011ch,Haskell:2017lkl,Kojo:2020ztt,Blaschke:2022egm,Yuan:2023dxl,Gholami:2024ety,Gholami:2025yqf,Yuan:2026yae}. Turning to the perturbative regime, the effects of a nonvanishing pairing gap were investigated in Ref.~\cite{Fukushima:2024gmp} through a unified treatment of pion superfluidity, two-color diquark superfluidity, and two-flavor color superconductivity within the Cornwall--Jackiw--Tomboulis formalism. Their results indicate that the speed of sound approaches the conformal limit from below and that the 2SC gap has only a modest impact on this quantity when compared with other physical mechanisms, a behavior attributed to the comparatively small magnitude of the gap. Nevertheless, the superconducting gap has a nontrivial impact on the behavior of the trace anomaly, seemingly allowing it to attain negative values. It is worth emphasizing that, so far, effective-model calculations generally predict that the speed of sound approaches the conformal limit from above~\cite{Gholami:2024ety,Azeredo:2026wgl,Andersen:2026sjc}. Nevertheless, the application of the NJL model in this regime necessarily involves an extrapolation beyond its natural domain of validity, under the assumption that cutoff artifacts have been properly controlled by the chosen regularization prescription.

Finally, we conclude this introduction by outlining the organization of the review. \mbox{Section~\ref{sec:model}} introduces the NJL model and the mean-field thermodynamic potential used throughout this work. In Section~\ref{sec:MFIR}, we review the magnetic-field-independent regularization procedure and discuss how it systematically removes magnetic-field-dependent divergences and associated artifacts. Section~\ref{sec:MSS} is devoted to the medium separation scheme, where we show how medium-dependent contributions can be consistently disentangled from vacuum divergences. The relevant thermodynamic quantities are presented in \mbox{Section~\ref{sec:thermo}}. Numerical applications are discussed in Section~\ref{sec:numerical}, where MFIR and MSS are combined to investigate dense magnetized quark matter and color superconductivity. Particular emphasis is given to the persistence of a nonvanishing superconducting gap at high densities and to the elimination of spurious effects generated by traditional regularization prescriptions. Finally, Section~\ref{sec:discussions} summarizes the main findings and discusses future perspectives.


\section{The NJL Model}
\label{sec:model}

In this work, we employ the SU(2)$_f$ NJL model, including both scalar-pseudoscalar and color-pairing interactions, to investigate magnetized quark matter. The effective Lagrangian density of the system in the presence of an external magnetic field is given by
\begin{eqnarray}
\mathcal{L} & = & \bar{q}\left[i\gamma^{\mu}\left(\partial_{\mu}-i\tilde{e}\tilde{\mathcal{Q}}\tilde{A}_{\mu}\right)+\hat{\mu}\gamma^{0}-\hat{m}\right]q
+G_{S}\left[\left(\bar{q}q\right)^{2}+\left(\bar{q}i\gamma_{5}\vec{\tau}q\right)^{2}\right]
\nonumber\\
 & + & G_{D}\left[\left(i\bar{q}^{C}\epsilon_{f}\epsilon_{c}^{3}\gamma_{5}q\right)\left(i\bar{q}\epsilon_{f}\epsilon_{c}^{3}\gamma_{5}q^{C}\right)\right].
 \label{LNJLCSC}
\end{eqnarray}

The quark fields $q=(q_u,q_d)^T$ are accompanied by their charge-conjugate counterparts, $q^C=C\bar q^T$ and $\bar q^C=q^TC$, where $C=i\gamma^2\gamma^0$ denotes the charge-conjugation matrix. Furthermore, $(\epsilon_c^3)^{ab}=(\epsilon_c)^{3ab}$ and $(\epsilon_f)^{ij}$ represent the antisymmetric tensors in color and flavor spaces, respectively. In Equation~(\ref{LNJLCSC}), $\hat m=\text{diag}(m_u,m_d)$ denotes the current quark mass matrix, while $\hat\mu=\{\mu_{fc}\}$ defines the corresponding chemical potentials. Throughout this work, we assume the isospin-symmetric limit, namely $m_u=m_d=m_c$ and $\mu_i=\mu$. In the presence of a color-superconducting gap $\Delta$, the coupling to the magnetic field $\tilde{\mathcal{A}}_{\mu}$ is defined through the rotated charge matrix
\begin{equation}
\tilde{Q} = \mathcal{Q}_f\otimes1_c - 1_f\otimes\left(\frac{\lambda_8}{2\sqrt{3}}\right),
\end{equation}
where $\mathcal{Q}_f=\text{diag}(2/3,-1/3)$ and $\lambda_8=\text{diag}(1,1,-2)/\sqrt{3}$ denotes the eighth Gell-Mann matrix in color space. This formulation ensures that a particular linear combination of the photon and gluon fields remains massless~\cite{Alford:1999pb,Allen:2015paa}, yielding the following rotated charges for the six flavor-color combinations:
\mbox{$q_{ur} = q_{ug} = 1/2,\;q_{ub} = 1,\;q_{dr} = q_{dg} = -1/2,\;q_{db} = 0$.}
 For a constant magnetic field in the $z$ direction ($\tilde{\mathcal{A}}_{\mu}=\delta_{\mu2}x_1B$), electromagnetic and color fields become mixed, leading to the rotated charge $\tilde e=e\cos(\theta)$, where $\theta$ denotes the mixing angle and is estimated to be of the order of $1/20$~\cite{Gorbar:2000ms}. For simplicity of notation, we will denote $\tilde{e}$ as $e$ throughout the remainder of this text.

 Within the mean-field approximation, the thermodynamic potential, $\Omega_T$, is given by
\begin{eqnarray}
\Omega_{T} & = & \frac{\left(M-m_{c}\right)^{2}}{4G_{S}}+\frac{\Delta^{2}}{4G_{D}}-\int\frac{d^{3}\vec{p}}{\left(2\pi\right)^{3}}\left[\xi\left(E_{p,0}^{+}\right)+\xi\left(E_{p,0}^{-}\right)\right]\nonumber \\
 &  & -\frac{eB}{8\pi^{2}}\sum_{n=0}^{\infty}\alpha_{n}\int\limits_{-\infty}^{+\infty}dp_{z}\bigg[\xi\left(E_{p,1}^{+}\right)+\xi\left(E_{p,1}^{-}\right) +2\xi\left(E_{p,\frac{1}{2}}^{+}\right)
 + 2\xi\left(E_{p,\frac{1}{2}}^{-}\right)\bigg],
 \label{OmegaT}
\end{eqnarray}
which explicitly incorporates the quantization of Landau levels induced by the external magnetic field $eB$\footnote{Here, $eB$ is given in GeV$^2$.}. 
In this expression, $\alpha_n = 2 - \delta_{n0}$ accounts for the degeneracy of the Landau levels, while the function $\xi$ is defined as
\begin{equation}
\xi(E_i^j) = E_i^j + 2T\ln\left(1 + e^{-E_i^j/T}\right),
\end{equation}
and $E_i^j$ denote the quasiparticle dispersion relations associated with the rotated charges $|a| = 0,1,1/2$, given by
\begin{align}
    &E^{\pm}_{p,0} = \sqrt{\vec{p}^2 + M^2} \pm \mu\,, \\
    &E^{\pm}_{p,1} = \sqrt{p_z^2 + 2 eBn + M^2} \pm \mu\,, \\
    &E^{\pm}_{p,\frac{1}{2}} = \sqrt{\left(\sqrt{p_z^2 + eBn + M^2} \pm \mu\right)^2 + \Delta^2 },
\end{align}
where $E_{p,a}=\sqrt{{\bf p}_{\perp,a}^{2}+p_{z}^{2}+M^{2}}$, while 
the transverse momentum is quantized according to
${\bf p}_{\perp,a}^{2}=2\left|a\right|eBn$ for $|a|\neq 0$
and
${\bf p}_{\perp,a}^{2}=p_{x}^{2}+p_{y}^{2}$ for $a = 0$.
Finally, the superscript labels the particle ($-\mu$) and antiparticle ($+\mu$) branches of the spectrum.

\vspace{0.2cm}

In the zero-temperature limit, the thermodynamic potential reduces to
\begin{eqnarray}
\Omega_{T = 0} &=& \frac{\left(M-m_{c}\right)^{2}}{4G_{S}}+\frac{\Delta^{2}}{4G_{D}} - \Omega_0 - \Omega_1 - \Omega_{\frac{1}{2}}
- \Omega_{\text{const}} \, ,
\label{OmegaT0}
\end{eqnarray}
where $\Omega_{\text{const}}$ denotes the vacuum contribution to the thermodynamic potential which is included to ensure that the pressure vanishes at $T = \mu = eB = 0$, and we have introduced the shorthand notation
\begin{eqnarray}
\Omega_0 & = & 2\int\frac{d^{3}\vec{p}}{\left(2\pi\right)^{3}}E_{p,0}+2\int\frac{d^{3}\vec{p}}{\left(2\pi\right)^{3}}(\mu - E_{p,0})\theta(\mu - E_{p,0}),\nonumber\\
\Omega_1 & = & \frac{eB}{2\pi^{2}}\sum_{n=0}^{\infty}\alpha_{n}\int\limits_{0}^{\infty}dp_{z}E_{p,1} + \frac{eB}{4\pi^{2}}\sum_{n=0}^{\infty}\alpha_{n}\int\limits_{0}^{\infty}dp_{z}(\mu - E_{p,1})\theta(\mu - E_{p,1}),\nonumber\\
\Omega_{\frac{1}{2}} & = & \frac{eB}{2\pi^{2}}\sum_{n=0}^{\infty}\alpha_{n}\int\limits_{0}^{\infty}dp_{z} \left(E_{p,\frac{1}{2}}^{+} + E_{p,\frac{1}{2}}^{-} \right).
\label{Omega_a}
\end{eqnarray}

In these expressions, the terms containing Heaviside functions are finite and emerge from the $T\rightarrow0$ limit of Equation~\eqref{OmegaT}, whereas the first integral appearing in each line corresponds to a vacuum contribution and is ultraviolet divergent, requiring regularization. In the following sections, we discuss the different regularization schemes employed to handle these divergences while avoiding the artificial regularization of medium-dependent contributions.

\section{Magnetic-Field-Independent Regularization---MFIR}
\label{sec:MFIR}
 Traditional regularization methods, such as sharp cutoffs $\theta(x - \Lambda)$ or smooth form factors~\cite{Fukushima:2007fc, Fayazbakhsh:2010bh, Fayazbakhsh:2010gc, Mandal:2012fq, Mandal:2016dzg, Mandal:2017ihr}, often introduce unphysical oscillations in the order parameters~\cite{Avancini:2019wed}.
These artifacts are frequently misinterpreted as genuine de Haas--van Alphen oscillations, particularly in regions where such oscillatory behavior should be absent~\cite{Allen:2015paa,Duarte:2015ppa,Duarte:2017nzv}. To address this issue, the magnetic-field-independent regularization procedure, which isolates divergent vacuum contributions from finite magnetic-field-induced effects, was originally proposed in Refs.~\cite{Ebert:1999ht,Ebert:2003yk}, subsequently applied in Ref.~\cite{Menezes:2008qt}, and later extended to include color superconductivity in Ref.~\cite{Allen:2015paa}. Since then, it has been widely discussed and applied in a variety of physical contexts.

Within the framework of this study, the explicit contributions corresponding to $|a| = 1$ and $|a| = 1/2$ to the total thermodynamic potential are evaluated.
Following the standard MFIR procedure (for comprehensive technical details see Refs.~\cite{Menezes:2008qt,Allen:2015paa}), the contribution associated with the $|a| = 1$ sector can be written as
\begin{eqnarray}
\Omega_0 & = & 4\int_{\Lambda}\frac{d^{3}\vec{p}}{(2\pi)^{3}}\bigg[\sqrt{\vec{p}^{2}+M^{2}} - \sqrt{\vec{p}^{2}+M_0^{2}}\bigg]
\nonumber\\
&+& 2\int\frac{d^{3}\vec{p}}{\left(2\pi\right)^{3}}\bigg[(\mu - E_{p,0})\theta(\mu - E_{p,0})\bigg],
\label{W0}
\end{eqnarray}
and
\begin{eqnarray}
\Omega_1 &=& \frac{(eB)^2}{2\pi^2} \left[\zeta'( -1, \chi) + \frac{\chi - \chi^2}{2} \ln(\chi) + \frac{\chi^2}{4} \right] \nonumber \\
&+& \frac{eB}{4\pi^2} \sum_{n=0}^{p_{B,\text{max}}} \alpha_n \Biggl[ \mu \sqrt{\mu^2 - p_B^2}   - p_B^2 \ln\bigg( \frac{\mu + \sqrt{\mu^2 - p_B^2}}{p_B} \bigg) \Biggr].
\label{W1}
\end{eqnarray}
Here, the quantities are defined as $\chi = \frac{M^2}{2eB}$, $p_{B}=\sqrt{M^{2}+2eBn}$, while the upper limit of the summation is given by $p_{B,\text{max}}=\left \lfloor \frac{\mu^{2}-M^{2}}{2eB} \right \rfloor$. It is worth noting that the prefactor of 4 appearing in the first integral of Equation~\eqref{W0} differs from the coefficient of 2 present in the original expression given in Equation~\eqref{Omega_a}. This modification originates from an algebraic rearrangement inherent to the MFIR framework, in which an additional contribution arising from the first term of $\Omega_1$ in Equation~\eqref{Omega_a} is absorbed into $\Omega_0$. Also, in the first line of Equation~\eqref{W0}, the subtracted dispersion relation with $M_0$ comes from $\Omega_{\text{const}}$ contribution in Equation~\eqref{OmegaT0}. This subtracted term also appears in $\Omega_{\frac{1}{2}}$ in the sequence. The thermodynamic contribution associated with the $|a| = 1/2$ sector, which incorporates the effects of diquark pairing, is given by
\begin{eqnarray}
\Omega_{\frac{1}{2}} &=& 4\int_{\Lambda} \frac{d^{3}\vec{p}}{(2\pi)^{3}} \left[ \sqrt{\left( \sqrt{\vec{p}^{2} + M^{2}} + \mu \right)^{2} + \Delta^{2}}  \right. \nonumber\\
&& \left. + \sqrt{\left( \sqrt{\vec{p}^{2} + M^{2}} - \mu \right)^{2} + \Delta^{2}}   -2 \sqrt{\vec{p}^{2}+M_0^{2}}\right]\nonumber \\
& & + \frac{eB}{4\pi^{2}} \int_{-\infty}^{+\infty} dp_{z} F\left(p_{z}^{2}\right) \nonumber \\
& & + \frac{(eB)^{2}}{2\pi^{2}} \left[ \zeta^{\prime}\left(-1, x\right) + \frac{(x - x^{2})}{2} \ln(x) + \frac{x^{2}}{4} \right] \nonumber \\
& & + \frac{eB}{2\pi^{2}} \int_{-\infty}^{+\infty} dp_{z} \left[\sum_{n=1}^{\infty} F\left(p_{z}^{2} + neB\right)
\right. \left. - \int_{0}^{\infty} dy~ F\left(p_{z}^{2} + eBy\right) \right], \label{W05}
\end{eqnarray}
where the dimensionless quantity $x = \frac{M^2 + \Delta^2}{eB}$ has been introduced. Finally, the auxiliary function $F(z)$ appearing in Equation~\eqref{W05} is defined by
\begin{eqnarray}
F(z)=\sqrt{\left(\sqrt{z+M^{2}}+\mu\right)^{2}+\Delta^{2}}
+\sqrt{\left(\sqrt{z+M^{2}} -\mu\right)^{2}+\Delta^{2}}
-2\sqrt{z+M^{2}+\Delta^{2}}.\label{Fz}
\end{eqnarray} 
More details about the analytical and numerical implementation of MFIR in this expression are given in the Appendix~\ref{appMFIR}.


\section{Medium Separation Scheme---MSS}
\label{sec:MSS}
In the presence of a superconducting gap, the quasiparticle dispersion relations take the form
\begin{align}
E_\Delta^\pm = \sqrt{(E_p \pm \mu)^2 + \Delta^2} \, ,
\end{align}
where $E_p = \sqrt{\vec{p}^2 + M^2}$, with $M$ denoting the effective quark mass, $\mu$ the quark chemical potential, and $\Delta$ the superconducting gap.
The loop contribution to the thermodynamic potential is then proportional to the momentum integral
\begin{align}
\int \frac{d^3 p}{(2\pi)^3} \left( E_\Delta^+ + E_\Delta^- \right) \, ,
\end{align}
which is ultraviolet divergent and therefore requires regularization.
We refer to the application of a regulator at this stage as the Traditional Regularization Scheme (TRS), which is commonly adopted in the literature.
It is important to notice, however, that this term depends explicitly on the quark chemical potential. Consequently, the regulator directly affects medium contributions, which should remain finite, rather than acting exclusively on the ultraviolet divergences originating from the vacuum sector.
A systematic way to disentangle these finite medium contributions from the divergent vacuum is provided by the medium separation scheme, which has proven to be a reliable framework for the description of a variety of phenomena, such as color superconductivity~\cite{Farias:2005cr,Duarte:2018kfd,Pasqualotto:2025kpo,Azeredo:2026wgl}, chiral imbalance~\cite{Farias:2016let,Das:2019crc,Azeredo:2024sqc,daSilva:2025koa,Zheng:2026jqs}, and pion superfluidity~\cite{Avancini:2019ego,Lopes:2021tro,Lopes:2025rvn}.

An overview of the method can be found in Ref.~\cite{XavierdeAzeredo:2026wlq}, where violations of the conformal bound at intermediate densities induced by superfluid and superconducting phenomena are discussed.
In particular, it is shown there that the MSS is crucial for a reliable description of the peak in the speed of sound, as well as for its convergence toward the conformal limit at high densities. Another argument supporting the MSS is the remarkable agreement of its predictions with those obtained from renormalizable quark--meson models~\cite{Andersen:2026xrf}, as well as from Renormalization Group (RG)-consistent~\cite{Gholami:2024diy} and RG-invariant mean-field~\cite{Brandt:2025tkg} formulations of the Nambu--Jona-Lasinio and quark--meson models, \mbox{respectively}. 
\par 
To illustrate the MSS prescription, let us consider the ultraviolet-divergent integrals present in the gap equations
\begin{align}
I_{\Delta} &= \sum_{s = \pm 1} \int \frac{d^3 p}{(2\pi)^3} \frac{1}{E_\Delta^s} \, , \\
I_{M,\frac{1}{2}} &= \sum_{s = \pm 1} \int \frac{d^3 p}{(2\pi)^3} \frac{1}{E_p} \frac{E_p + s\mu}{E_\Delta^s} \, , 
\end{align}
where $s=\pm 1$. Notice that, although these terms no longer depend on the magnetic field after the MFIR procedure is applied, they remain explicitly $\mu$-dependent. We begin by introducing an integral over an auxiliary variable, analogous to the fourth component of momentum in Euclidean space:
\begin{align}\label{eqn:mss_p4integral}
\int \frac{d^3 p}{(2\pi)^3} \frac{1}{\sqrt{(E_p + s\mu)^2 + \Delta^2}} = \int\frac{d^3 p}{\left(2\pi\right)^{3}}\left[\frac{1}{\pi}\int_{-\infty}^{+\infty}\frac{da}{a^2 + (E_p + s\mu)^2 + \Delta^2}\right].
\end{align}
We then make use of the identity
\begin{align}
\frac{1}{a^2 + (E_p + s \mu)^2 + \Delta^2} &= \frac{1}{a^2 + \vec{p}^2 + M_0^2} + \frac{M_0^2 - (M^2 + \Delta^2 + \mu^2 + 2 s \mu E_p)}{\left(a^2 + \vec{p}^2 + M_0^2\right) \left[a^2 + (E_p + s\mu)^2 + \Delta^2\right]} \, ,
\end{align}
which can be iterated since the left-hand side also appears in the last term on the right-hand side. After two iterations, one obtains
\begin{align}\label{eqn:mss_alliterations}
\frac{1}{a^2 + \left(E_p + s\mu\right)^2 + \Delta^2} &= \frac{1}{a^2 + \vec{p}^2 + M_0^2} + \frac{A - 2 s \mu E_p}{\left(a^2 + \vec{p}^2 + M_0^2\right)^2} + \frac{\left(A - 2 s \mu E_p \right)^2}{\left(a^2 + \vec{p}^2 + M_0^2\right)^3} \nonumber \\
&+ \frac{\left(A - 2 s \mu E_p \right)^3}{\left(a^2 +
\vec{p}^2 + M_0^2\right)^3 \left[a^2 + \left(E_p + s\mu\right)^2 + \Delta^2\right]}.
\end{align}
The key point is that the scale of the medium separation scheme is set by $M_0 = M(T = \mu = eB = 0)$, namely the effective quark mass evaluated in vacuum and therefore unaffected by medium contributions. For clarity, we have also introduced the auxiliary variable $A = M_0^2 - M^2 - \Delta^2 - \mu^2$.
Substituting this expression into the integral $I_{\Delta}$, the sum over $s$ and the integration over $a$ can be carried out analytically by using the residue theorem. For the last term in Equation~\eqref{eqn:mss_alliterations}, we keep the sum explicit and make use of the Feynman parametrization
\begin{align}
\frac{1}{A_1^3 A_2} = 3 \int_{0}^{1} dt \, \frac{(1 - t)^2}{\left[ (A_2 - A_1)t + A_1 \right]^4} \, .
\end{align}
The resulting expression is

\begin{align}
I_\Delta^{\rm MSS} &= \sum_{s = \pm 1} \int\frac{d^3 p}{(2\pi)^3} \frac{1}{\pi} \int_{-\infty}^{+\infty} \frac{da}{a^2 + (E_\Delta^s)^2} \nonumber\\
&= 2 I_{\rm quad}(M_0) + \left( M_0^2 - M^2 - \Delta^2 + 2\mu^2 \right) I_{\rm log}(M_0) \nonumber \\
& + \frac{3}{4} \left[ A^2 + 4 \mu^2 (M^2 - M_0^2) \right] I_1 + 2 I_2 \, , 
\label{idd05}
\end{align}
where we have defined
\begin{align}
&I_{\rm quad} (M_0) = \int \frac{d^3 p}{(2\pi)^3} \frac{1}{\sqrt{\vec{p}^2 + M_0^2}} \, , \nonumber \\
&I_{\rm log} (M_0) = \int \frac{d^3 p}{(2\pi)^3} \frac{1}{(\vec{p}^2 + M_0^2)^{3/2}} \, , \nonumber \\
&I_1 = \int \frac{d^3 p}{(2\pi)^3} \frac{1}{(\vec{p}^2 + M_0^2)^{5/2}} \, , \nonumber \\
&I_2 = \frac{15}{32} \sum_{s = \pm 1} \int \frac{d^3 p}{(2\pi)^3} \int_{0}^{1} dt \, (1 - t)^2 \frac{(A - 2 s \mu E_p)^3}{\left[ \vec{p}^2 + M_0^2 + (2 s \mu E_p - A) t \right]^{7/2}} \, .
\end{align}
Only the integrals $I_{\rm quad}(M_0)$ and $I_{\rm log}(M_0)$ are ultraviolet divergent, and they depend exclusively on the effective mass in vacuum.
Both $I_1$ and $I_2$ can be integrated over the entire momentum space, and only the latter carries medium-dependent contributions.
We now turn to the gap equation for the effective quark mass.
One may immediately notice~that
\begin{align}
I_{M,\frac{1}{2}}^{\rm MSS} &= \sum_{s = \pm 1} \int \frac{d^3 p}{(2\pi)^3} \frac{1}{E_p} \frac{E_p + s \mu}{E_\Delta^s} \nonumber \\
&= I_{\Delta}^{\rm MSS} + \sum_{s = \pm 1} \int \frac{d^3 p}{(2\pi)^3} \frac{s \mu}{E_p} \frac{1}{E_\Delta^s} \, .
\end{align}
The evaluation of the remaining integral follows essentially the same steps as before. We introduce an auxiliary integral, as in Equation~\eqref{eqn:mss_p4integral},
\begin{align}
\int \frac{d^3 p}{(2\pi)^3} \frac{s\mu}{E_p} \frac{1}{\sqrt{(E_p + s\mu)^2 + \Delta^2}} = \int \frac{d^3 p}{(2\pi)^3} \frac{s\mu}{E_p} \left[ \frac{1}{\pi} \int_{-\infty}^{+\infty} \frac{da}{a^2 + (E_p + s \mu)^2 + \Delta^2} \right] \, ,
\end{align}
before applying the identity given in Equation~\eqref{eqn:mss_alliterations}.
The main difference in this case is that the multiplicative factor $s$ changes which terms remain after the summation is performed, according to their parity.
With these considerations, one obtains
\begin{align}
I_{M,\frac{1}{2}}^{\rm MSS} &= 2 I_{\rm quad}(M_0) + \left( M_0^2 - M^2 - \Delta^2\right) I_{\rm log}(M_0) \nonumber \\
& + \frac{3}{4} \left[ A^2 + 4 \mu^2 (M^2 - M_0^2 - A) \right] I_1 + 2 I_2 + I_4 \, , 
\label{Imd05}
\end{align}
where
\begin{align}
I_4 = \frac{15}{16} \sum_{s = \pm 1} \int \frac{d^3 p}{(2\pi)^3} \int_{0}^{1} dt \, (1 - t)^2 \, \frac{s \mu}{E_p} \, \frac{(A - 2 s \mu E_p)^3}{\left[ \vec{p}^2 + M_0^2 + ( 2 s \mu E_p - A) t \right]^{7/2}} \, .
\end{align}

In the absence of a superconducting gap, the effective-mass equation takes a simpler form involving the integral
\begin{align}
I_{M,0} &= \int \frac{d^3 p}{(2\pi)^3} \frac{1}{E_p} \, .
\end{align}
Although this expression is not explicitly dependent on $\mu$, it still carries an implicit medium dependence through the dynamical quark mass $M$.
For consistency, we rewrite this expression within the MSS framework so that only vacuum quantities are affected by the regulator.
Once again, we introduce an auxiliary integral, yielding
\begin{align}
I_{M,0} &= \int \frac{d^3 p}{(2\pi)^3} \frac{1}{\pi} \int_{-\infty}^{+\infty} da \frac{1}{a^2 + \vec{p}^2 + M^2} \, .
\end{align}
Next, we employ the identity
\begin{align}
\frac{1}{a^2 + \vec{p}^2 + M^2} &= \frac{1}{a^2 + \vec{p}^2 + M_0^2} + \frac{M_0^2 - M^2}{\left(a^2 + \vec{p}^2 + M_0^2\right)^2} \nonumber \\
&+ \frac{(M_0^2 - M^2)^2}{\left(a^2 + \vec{p}^2 + M_0^2\right)^2 \left(a^2 + \vec{p}^2 + M^2\right) } \, ,
\end{align}
requiring one fewer iteration than in the previous two cases, since the last term is already finite when integrated over the entire momentum space.
The integration over $a$ can be performed analytically, while the last term is treated through the Feynman parametrization
\begin{align}
\frac{1}{A_1^2 A_2} = 2 \int_{0}^{1} dt \, \frac{(1-t)}{\left[ (A_2 - A_1) t + A_1 \right]^3} \, .
\end{align}
Finally, one obtains
\begin{align}
I_{M,0}^{\rm MSS} &= I_{\rm quad}(M_0) + \frac{M_0^2 - M^2}{2} I_{\rm log}(M_0) + I_3 \, ,
\label{Imd0}
\end{align}
where
\begin{align}
I_3 &= \frac{3}{4} \int \frac{d^3 p}{(2\pi)^3} \int_{0}^{1} dt \, (1 - t) \, \frac{(M_0^2 - M^2)^2}{\left[ \vec{p}^2 + M_0^2 + (M^2 - M_0^2) t \right]^{5/2}} \, .
\end{align}
The MSS-regularized thermodynamic potential can be obtained by integrating the corresponding gap equations.
It contains the contribution
\begin{align}
I_\Omega &= 4 \int \frac{d^3 p}{(2\pi)^3} \left(E_p + E_\Delta^- + E_\Delta^+ - 3 E_0 \right) \nonumber \\
&= 4 \int \frac{d^3 p}{(2\pi)^3} \left(E_p - E_0\right) + 4 \sum_{s = \pm 1} \int \frac{d^3 p}{(2\pi)^3} \left(E_\Delta^s - E_0\right) \, ,
\end{align}
and its final form becomes
\vspace{-6pt}%
\begin{align}
I_\Omega^\text{MSS} &= 2 (M^2 - M_0^2 + 2 \bar{M}) I_{\rm quad} (M_0) - \left[ \frac{(M^2 - M_0^2)^2}{2} + \bar{M}^2 - 4 \Delta^2 \mu^2 \right] I_{\rm log} (M_0) \nonumber \\
&+ 4 \int \frac{d^3 p}{(2\pi)^3} \left[ \frac{(M^2 - M_0^2)^2 + 2 \bar{M}^2 - 8 \Delta^2 \mu^2}{8 E_0^3} - \frac{M^2 - M_0^2 + 2\bar{M}}{2 E_0} + E_p + E_\Delta^+ + E_\Delta^- - 3 E_0 \right] \, ,
\end{align}

where $E_0 = \sqrt{\vec{p}^2 + M_0^2}$ and $\bar{M} = \Delta^2 + M^2 - M_0^2$.

\par
\begin{figure}[htpb!]
\centering 
{\includegraphics[scale=0.3]{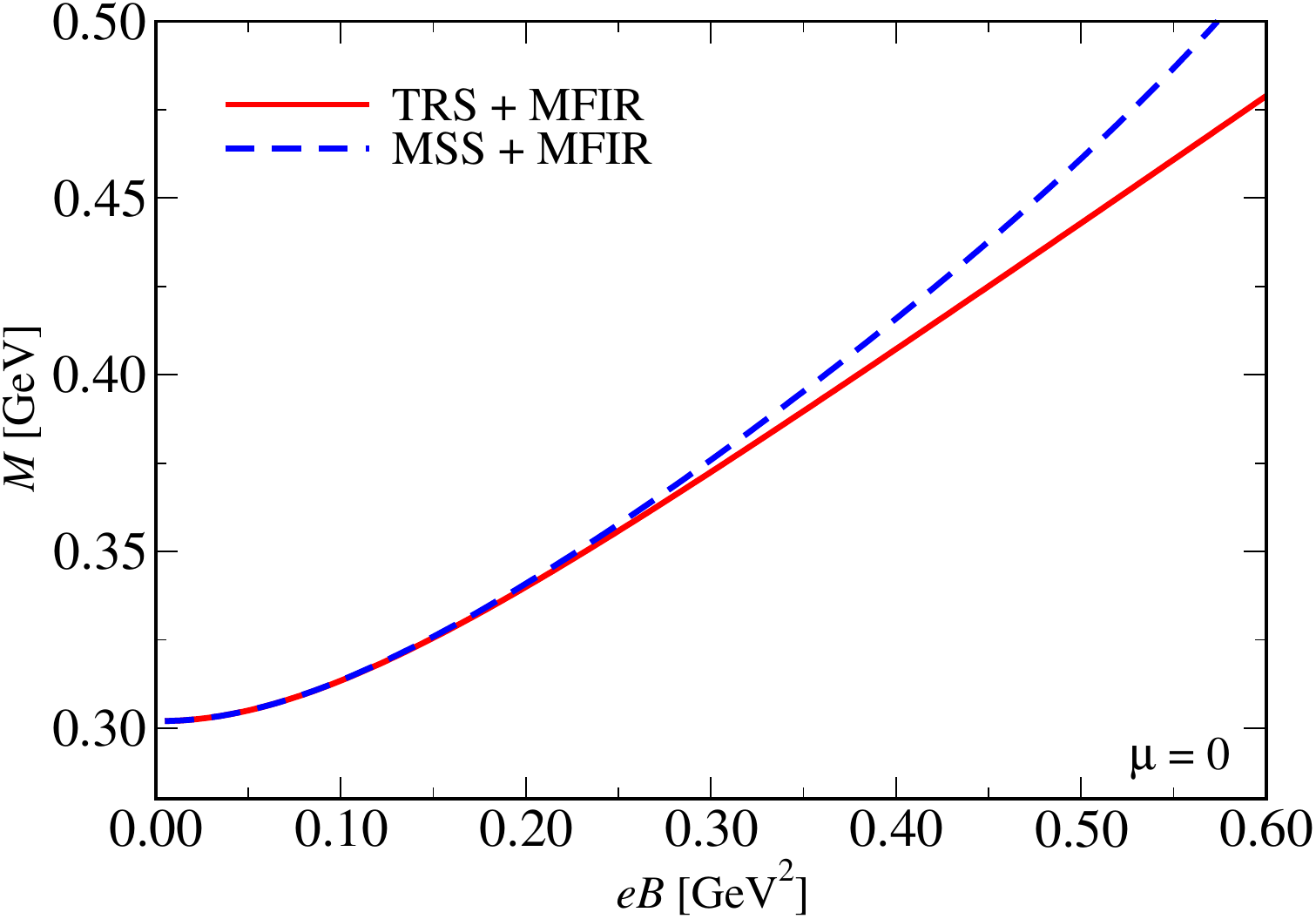}}
{\includegraphics[scale=0.3]{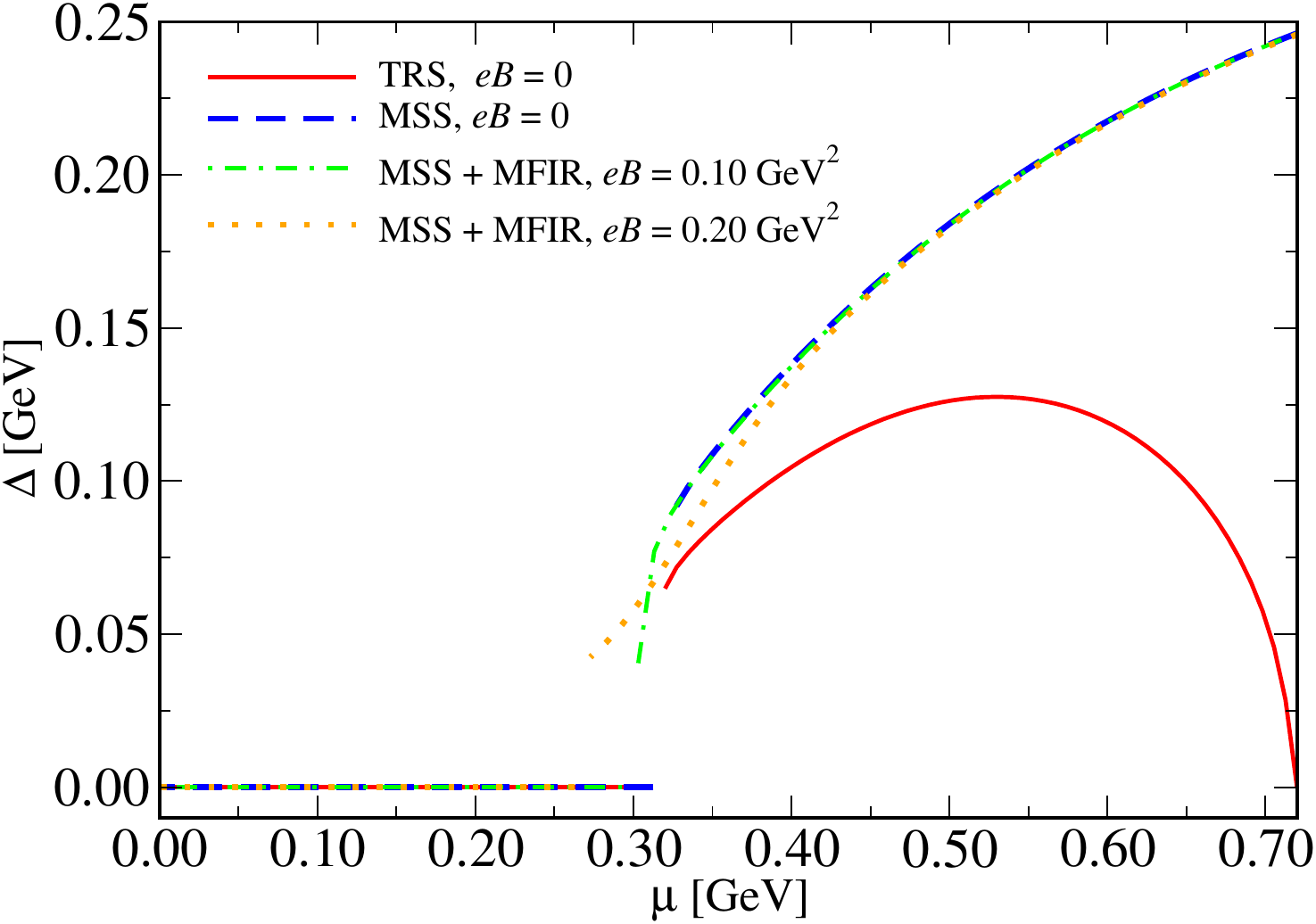}}
  \caption{Effective quark mass $M$ as a function of the magnetic field $eB$ for $\mu = 0$ (\textbf{left}
 panel), and diquark condensate $\Delta$ as a function of the chemical potential $\mu$ for different values of $eB$ (\textbf{right} panel).}
   \label{Fig1}
  \end{figure}

 The left panel of Figure~\ref{Fig1} illustrates the effective quark mass, $M$, as a function of the magnetic field strength, $eB$, at vanishing chemical potential.
As can be seen, the quark mass increases monotonically with the magnetic field for both regularization schemes, exhibiting the well-known phenomenon of magnetic catalysis~\cite{Menezes:2008qt,Shovkovy:2012zn,Ramos:2013aia}.
More specifically, at low magnetic fields, the results obtained from the TRS + MFIR and MSS + MFIR schemes are nearly identical.
However, as the magnetic field strength increases, the two curves gradually separate, with the MSS + MFIR scheme yielding a noticeably larger quark mass than the TRS + MFIR approach at $eB = 0.60~\mathrm{GeV}^2$.
The right panel displays the diquark condensate $\Delta$ as a function of the chemical potential $\mu$ for different values of the magnetic field $eB$.
An important feature emerges when comparing the regularization schemes: within the TRS framework at $eB = 0$, $\Delta$ reaches a maximum value before rapidly dropping to zero as $\mu$ increases. This behavior is clearly unphysical and contrasts with recent results available in the literature obtained from alternative approaches~\cite{Son:1998uk, Kogut:2000ek, Kogut:2001na, Braguta:2016cpw, Gholami:2024diy,Andersen:2025ezj,Andersen:2026xrf,Andersen:2026sjc}. A particularly important example is the persistence of the superconducting gap at high densities, a feature that is not reproduced within the TRS framework, as demonstrated in the right panel of Figure~\ref{Fig1}. The same qualitative behavior is observed for finite values of $eB$. Conversely, within the MSS framework, $\Delta$ increases monotonically with $\mu$ for both vanishing and nonvanishing values of $eB$.
Furthermore, the introduction of a magnetic field within the MSS + MFIR scheme slightly shifts the onset of the superconducting phase toward lower values of $\mu$, while the qualitative behavior of $\Delta$ at higher chemical potentials remains robust and nearly independent of $eB$ throughout the entire range of magnetic fields considered.

\section{Thermodynamic Quantities}
\label{sec:thermo}
In this section, we introduce the thermodynamic observables required to characterize the magnetized quark matter system.
In the presence of an external magnetic field, the pressure along the direction of the field, $p$, is obtained directly from the grand canonical potential evaluated at zero temperature, Equation~\eqref{OmegaT0}, through the standard thermodynamic relation\footnote{It is worth noting that a strong magnetic field naturally induces an anisotropy in the system, splitting the thermodynamic pressure and the speed of sound into distinct components parallel and perpendicular to the field direction~\cite{Ferrer:2010wz,Ferrer:2020tlz,Ferrer:2022afu}. A detailed analysis of this spatial anisotropy within the current framework was performed in Ref.~\cite{Azeredo:2026wgl}. In the present work, however, we restrict our focus to the longitudinal quantities for simplicity.}:
\begin{equation}
p(M,\mu,eB) = -\Omega_{T=0}(M,\mu, eB).
\label{Pressure}
\end{equation}

To ensure a physical normalization in which the pressure vanishes in the vacuum state ($\mu = 0$), we subtract the vacuum energy contribution and define the normalized pressure $p_N$ as
\begin{equation}
p_N = p(M,\mu,eB) - p(M_{0,B},0,eB),
\label{PparN}
\end{equation}
where $M_{0,B}$ denotes the constituent quark mass evaluated at $\mu = 0$, but finite $eB$. For brevity, explicit functional dependencies will be omitted hereafter. Note that the pure magnetic field contribution, $-B^2/2$, is absent from Equation~\eqref{Pressure}, since it cancels naturally through the regularization procedure defined in Equation~\eqref{PparN}. The system's magnetization, \text{{\termchancery M}}, which quantifies the response of the medium to variations in the external magnetic field, is obtained from the derivative of the normalized pressure at fixed chemical potential:
\begin{equation}
\text{{\termchancery M} }  = \left.\frac{\partial p_N}{\partial B}\right|_{\mu}.
\label{magnetization}
\end{equation}
Similarly, the baryon number density $n_B$ is defined as the response to variations in the baryon chemical potential, $\mu_B = N_c\mu$, at fixed magnetic field:
\begin{equation}
n_B = \left.\frac{\partial p_N}{\partial \mu_B}\right|_{eB}.
\label{rho}
\end{equation}
Using these quantities, the normalized energy density $\varepsilon_N$ can be obtained through the standard Euler relation for magnetized media:
\begin{equation}
\varepsilon_N = -p_N + \mu_B n_B +\text{{\termchancery M} } B.
\label{epsilon}
\end{equation}
Finally, the longitudinal speed of sound, $c_s^2$, which characterizes the propagation of perturbations along the magnetic-field direction, is determined at fixed field strength through
\begin{equation}
c_s^2 = \left.\frac{d p_N}{d \varepsilon_N}\right|_{eB}.
\label{cs2parallel}
\end{equation}

\section{Numerical Results }
\label{sec:numerical}

To obtain the numerical results, we must first specify the model parameters:
the current quark mass $m_c$, the scalar coupling $G_S$, and the ultraviolet scale $\Lambda$. These parameters are calibrated using empirical inputs, namely the pion decay constant $f_{\pi} = 93.2$ MeV, the pion mass $m_{\pi} = 135$ MeV, and the vacuum quark condensate $\langle \bar{q} q \rangle ^{1/3}_0 = -250$ MeV.
From this fitting procedure, we obtain $G_S = 4.75$ GeV$^{-2}$, $m_c = 4.99$ MeV, and $\Lambda = 660$ MeV. Across all regularization schemes, the resulting vacuum quark mass, which also defines the characteristic mass scale of the MSS framework, is approximately $M_0 \sim 302$ MeV. The diquark coupling constant $G_D$ may be fixed by a Fierz transformation, $G_D = 0.75G_S$~\cite{Buballa:2003qv}, or treated as a free parameter through the ratio $\eta\equiv G_D/G_S$.

\begin{figure}[htpb!]
  \centering
  {\includegraphics[scale=0.3]{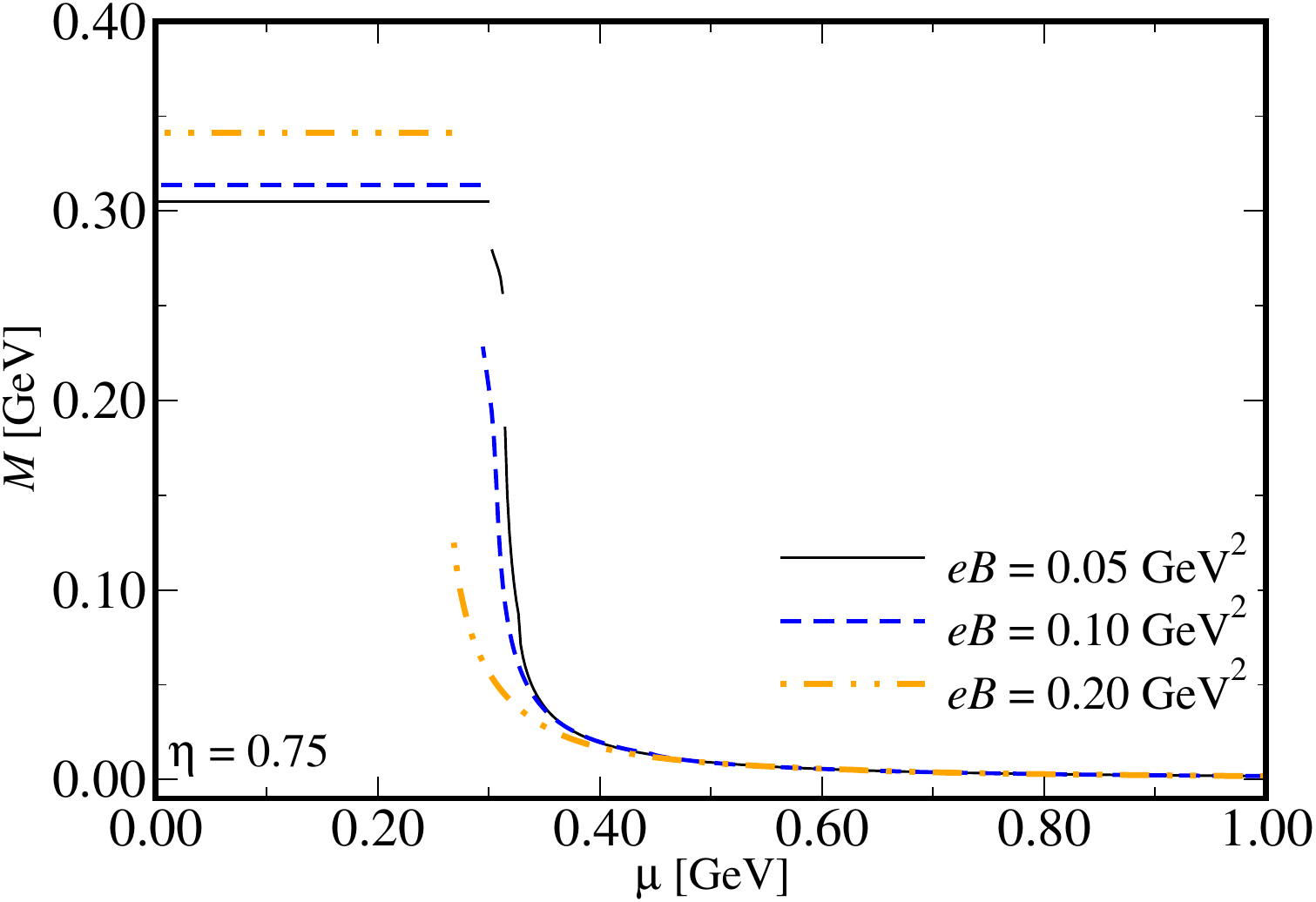}}
{\includegraphics[scale=0.3]{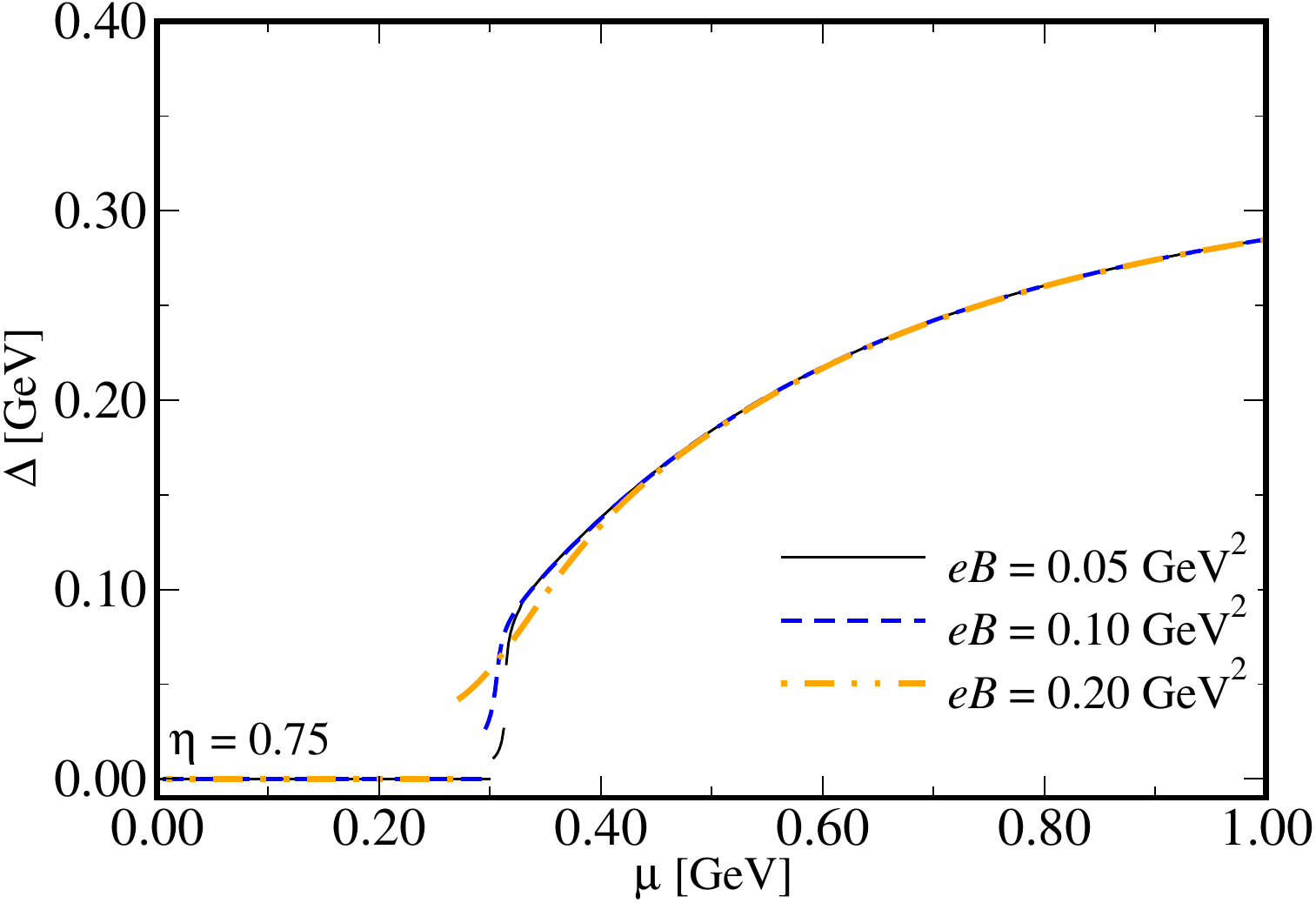}}\\
{\includegraphics[scale=0.3]{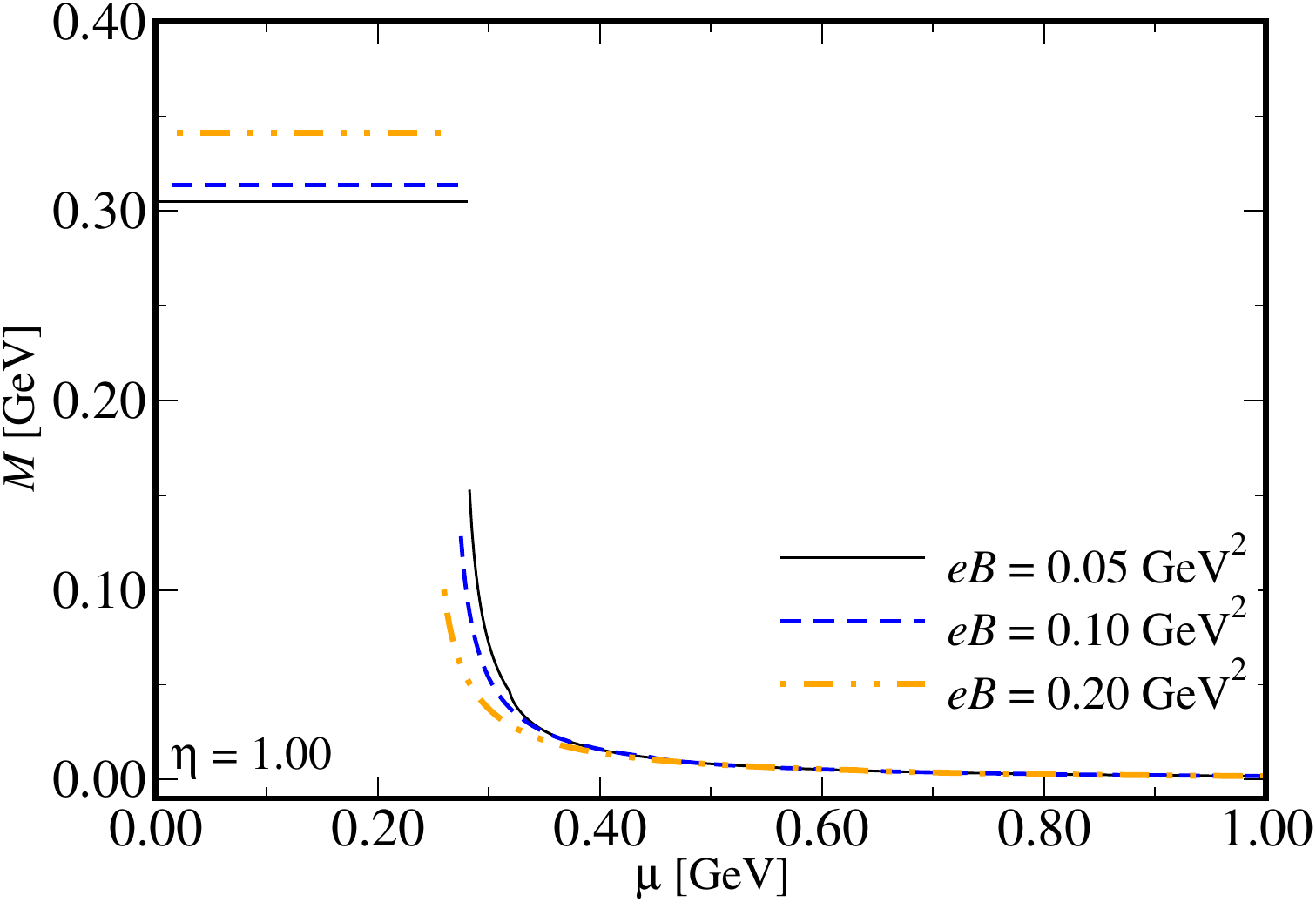}}
{\includegraphics[scale=0.3]{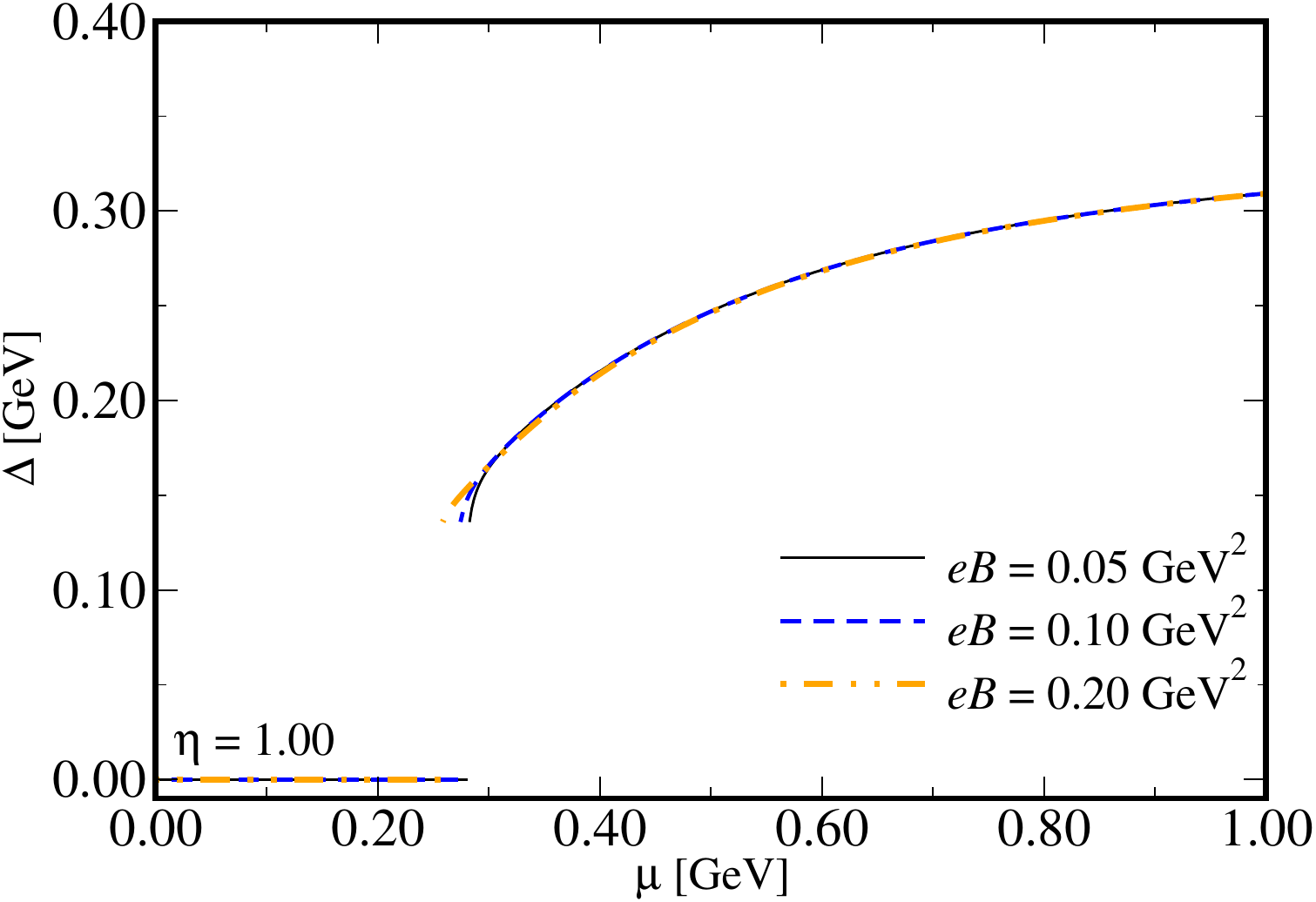}}\\
{\includegraphics[scale=0.3]{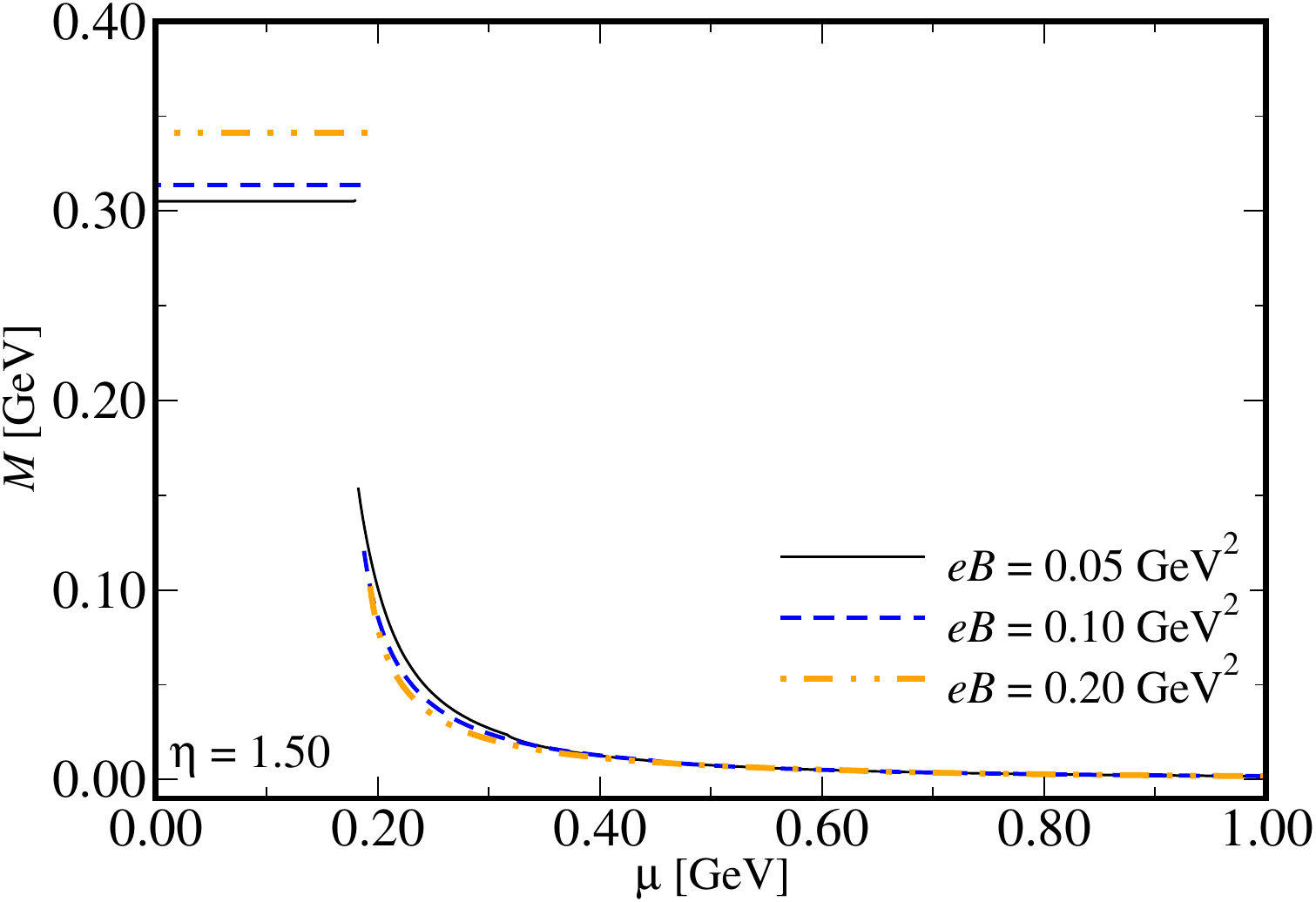}}
{\includegraphics[scale=0.3]{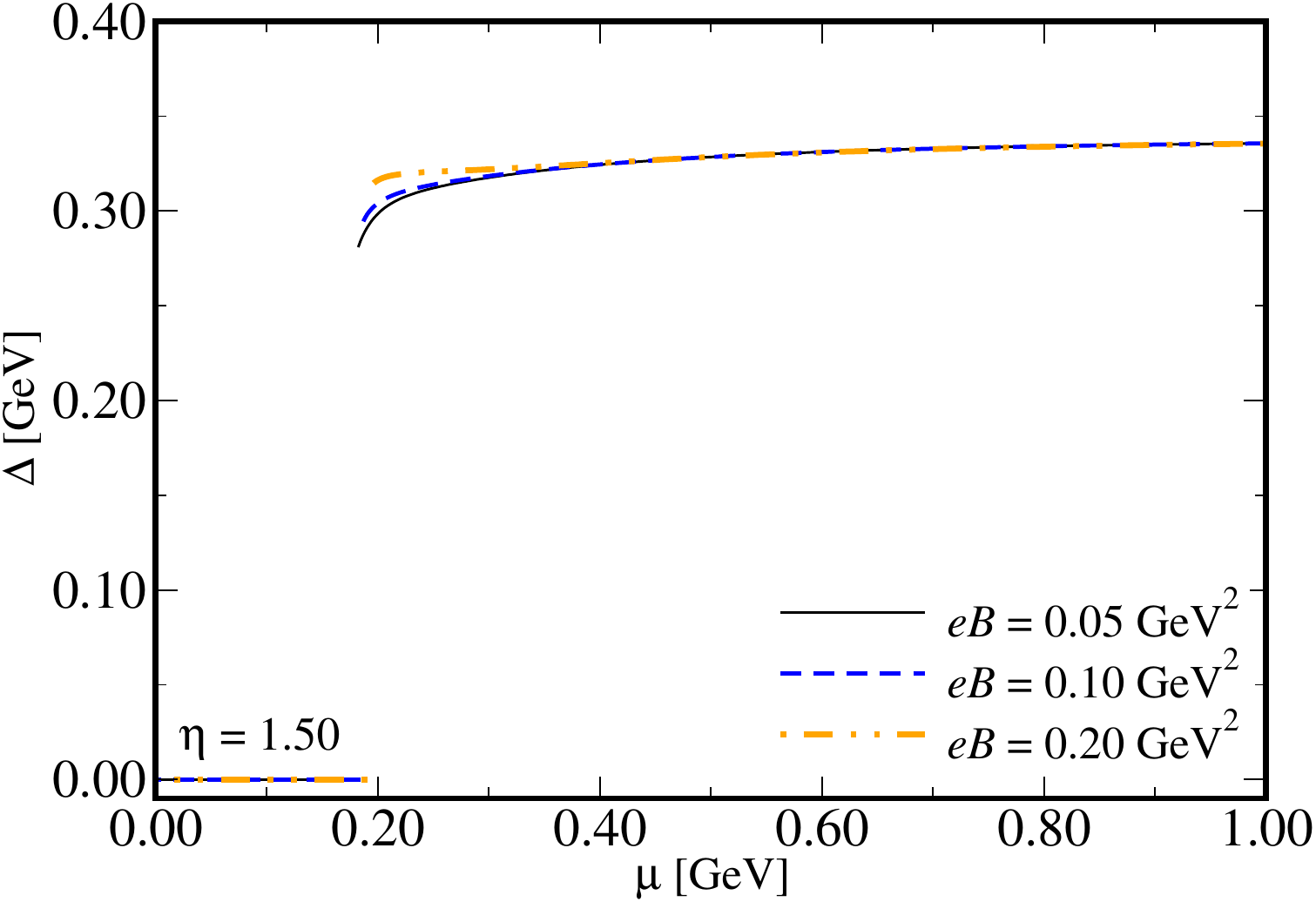}}
  \caption{Constituent quark mass $M$ (\textbf{left} panels) and diquark condensate $\Delta$ (\textbf{right} panels) as a function of the chemical potential $\mu$ within the MSS + MFIR method at finite magnetic field. The results are presented for three different values of the ratio between couplings $\eta$.}
  \label{Fig2}
\end{figure}

Figure~\ref{Fig2} displays the constituent quark mass $M$ (left panels) and the diquark condensate $\Delta$ (right panels) as functions of the chemical potential $\mu$ within the MSS + MFIR framework at finite magnetic field.
As the coupling ratio $\eta$ increases, the first-order character of the phase transition becomes more pronounced.
Conversely, for the lowest analyzed coupling ratio ($\eta = 0.75$) combined with a magnetic field of $eB = 0.05\text{ GeV}^2$, the system exhibits a relatively weak first-order phase transition.
Furthermore, it is worth noting that within the MSS framework, the diquark condensate consistently behaves as a monotonically increasing function of the chemical potential in the high-density regime for all values of $\eta$ considered. This behavior was also verified for alternative vacuum parametrizations that lead to different values of $M_0$.
Regarding the influence of the external magnetic field, a clear shift of the critical chemical potential $\mu_c$ is observed in all panels.
Specifically, as the magnetic field strength $eB$ increases from $0.05\text{ GeV}^2$ to $0.20\text{ GeV}^2$, the location of the first-order phase transition systematically shifts toward lower values of the chemical potential.
This behavior indicates that a stronger magnetic field accelerates the partial restoration of chiral symmetry and the subsequent onset of diquark condensation, thereby reducing the critical energy density required for the phase transition to occur.

\begin{figure}[htpb!] 
\centering
{\includegraphics[scale=0.3]{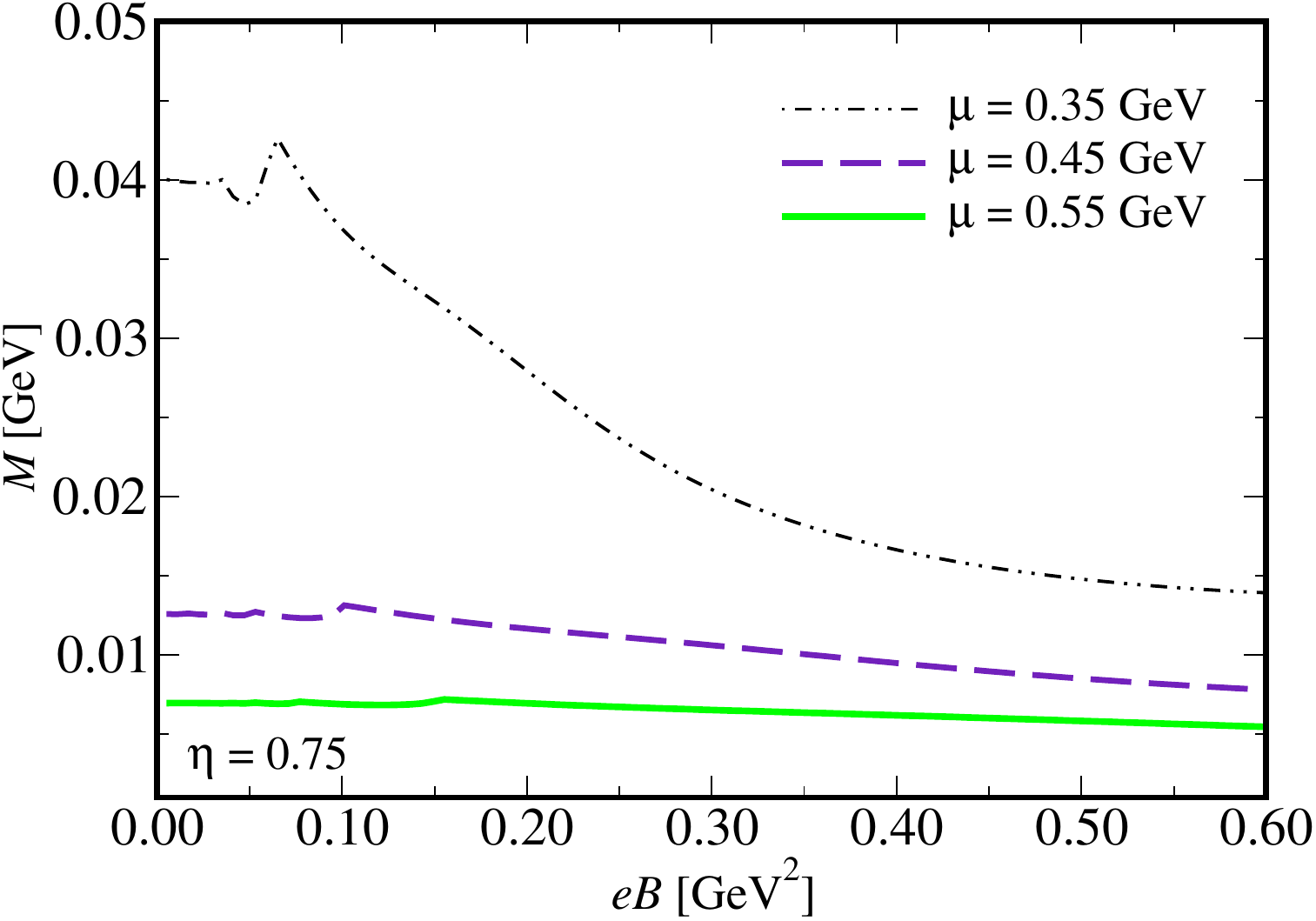}}
{\includegraphics[scale=0.3]{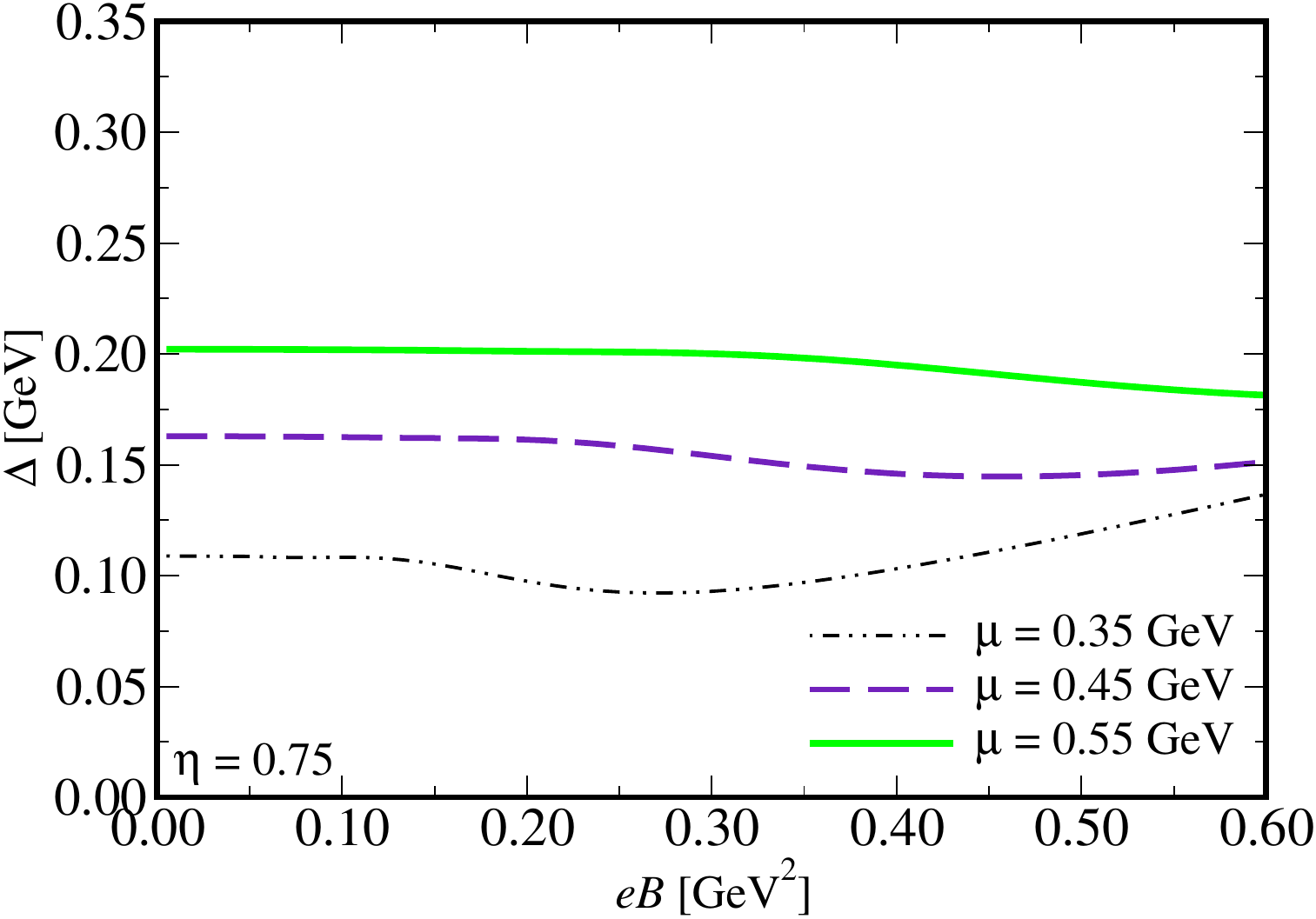}}\\
{\includegraphics[scale=0.3]{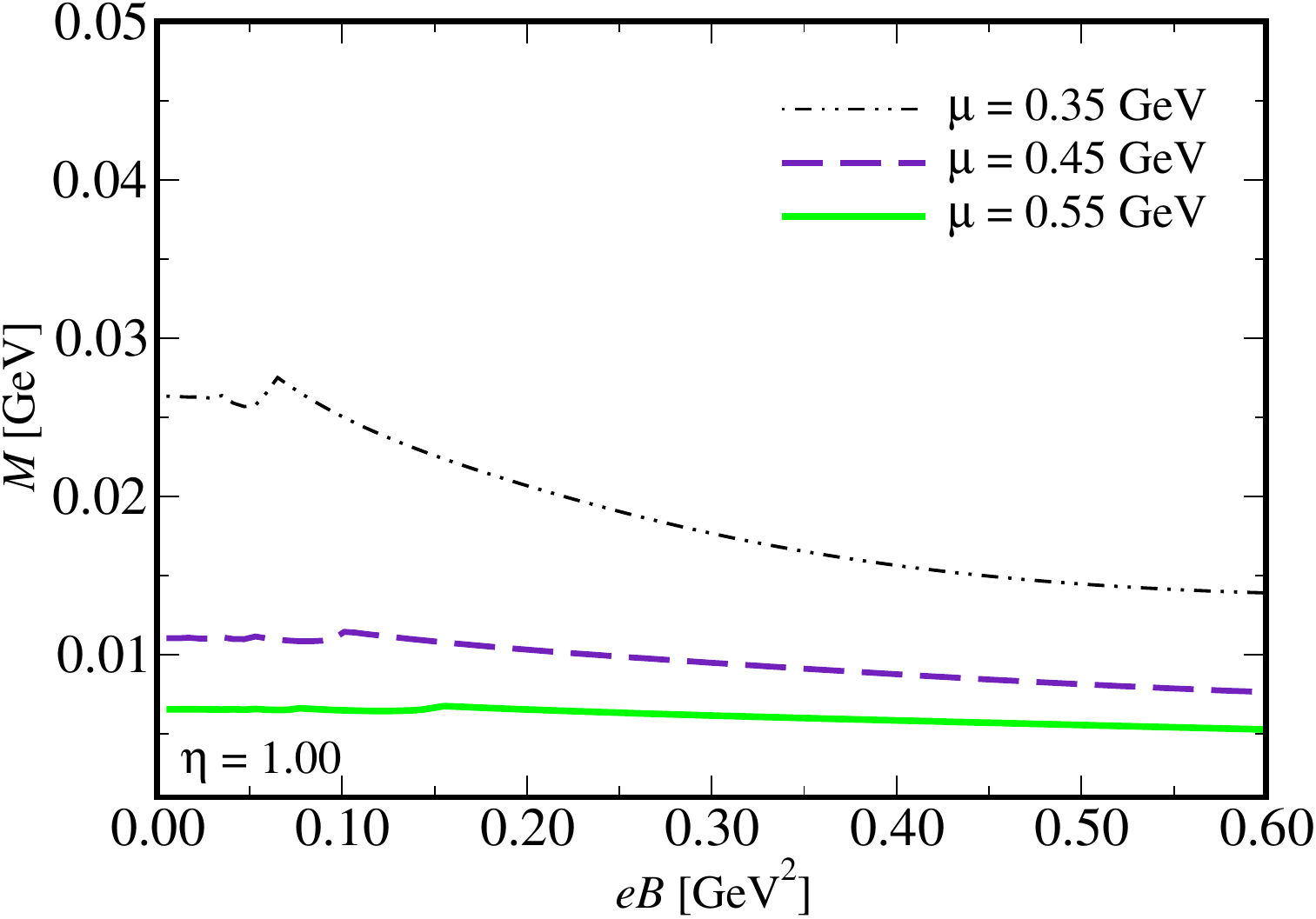}}
{\includegraphics[scale=0.3]{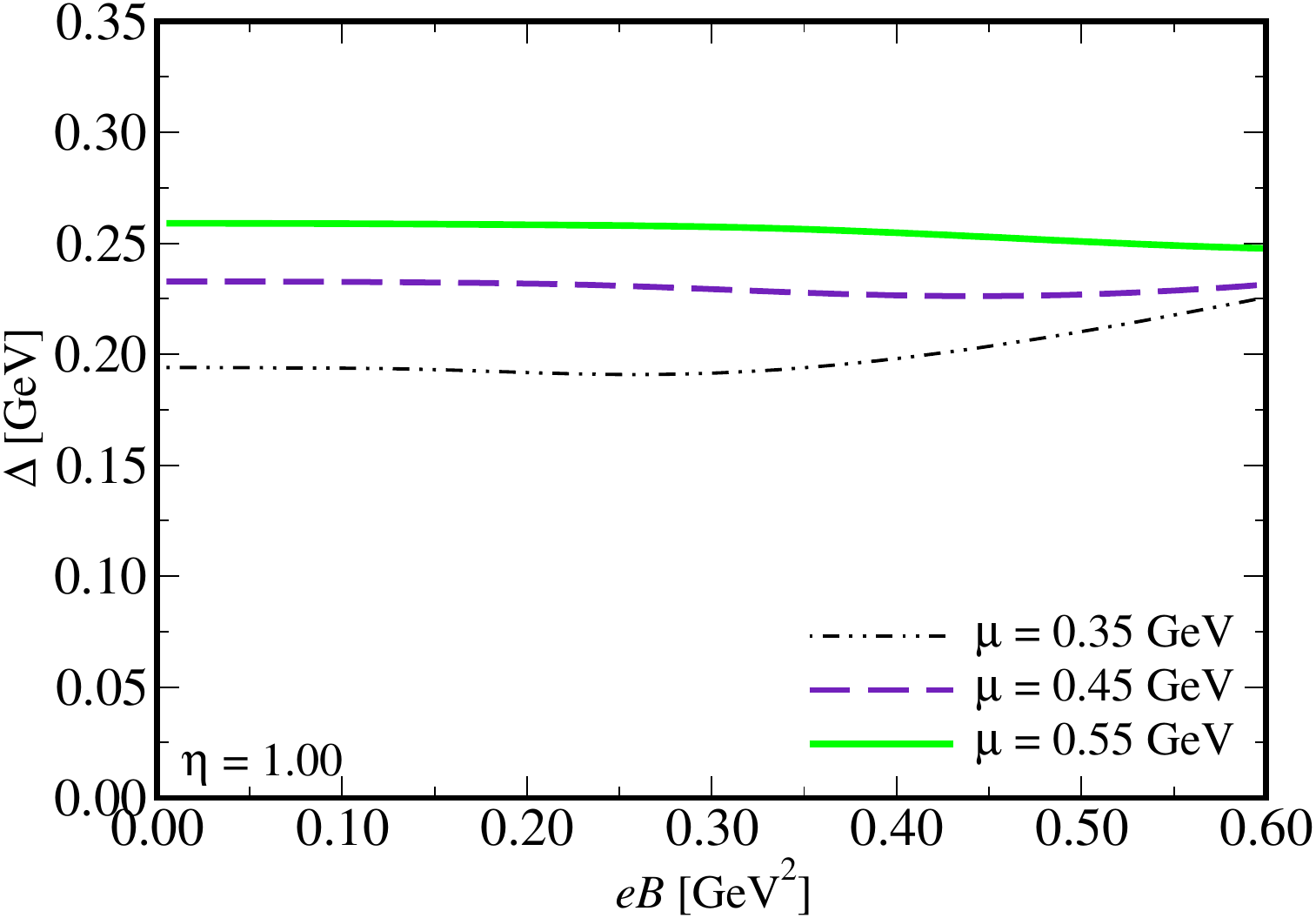}}\\
{\includegraphics[scale=0.3]{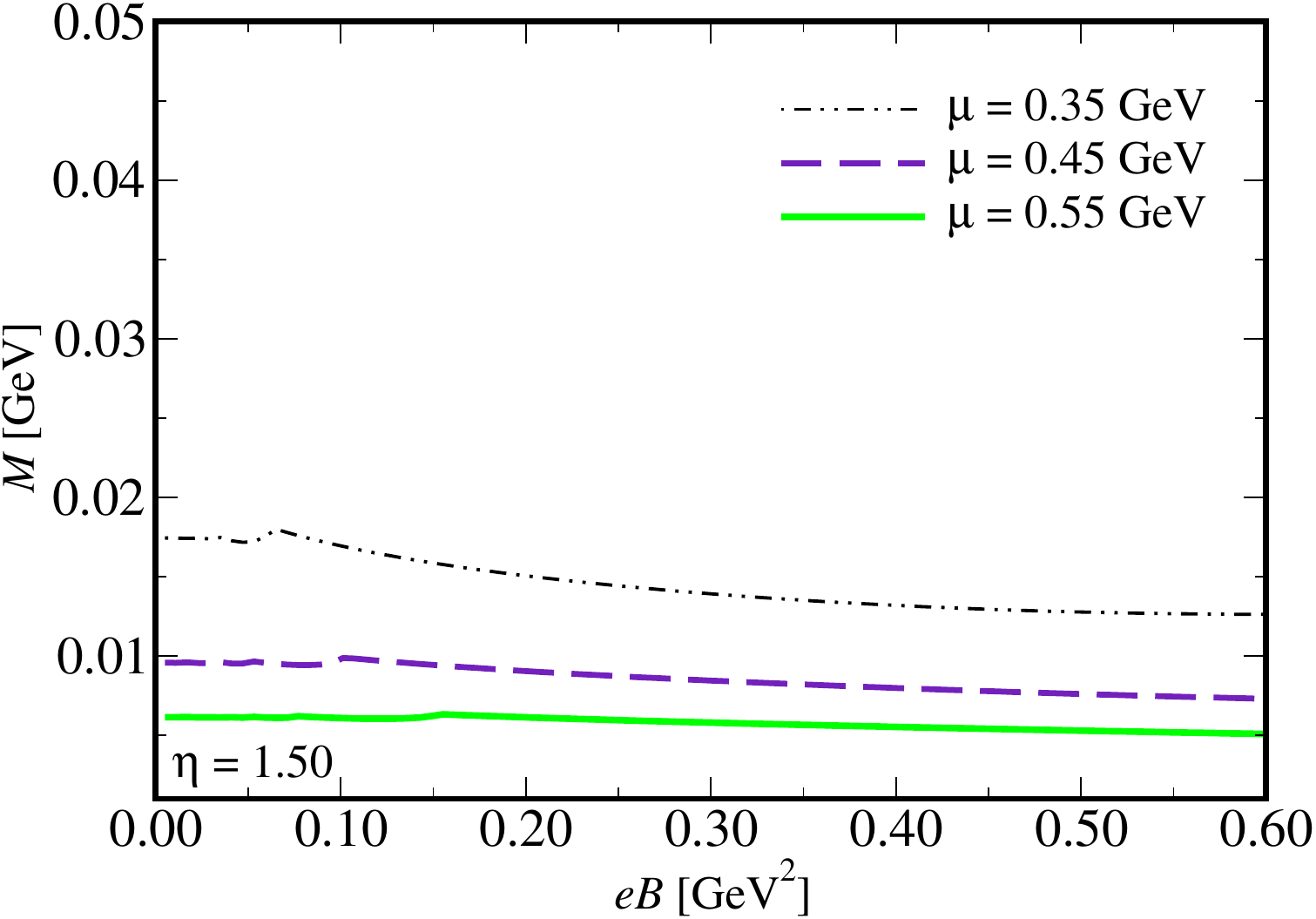}}
{\includegraphics[scale=0.3]{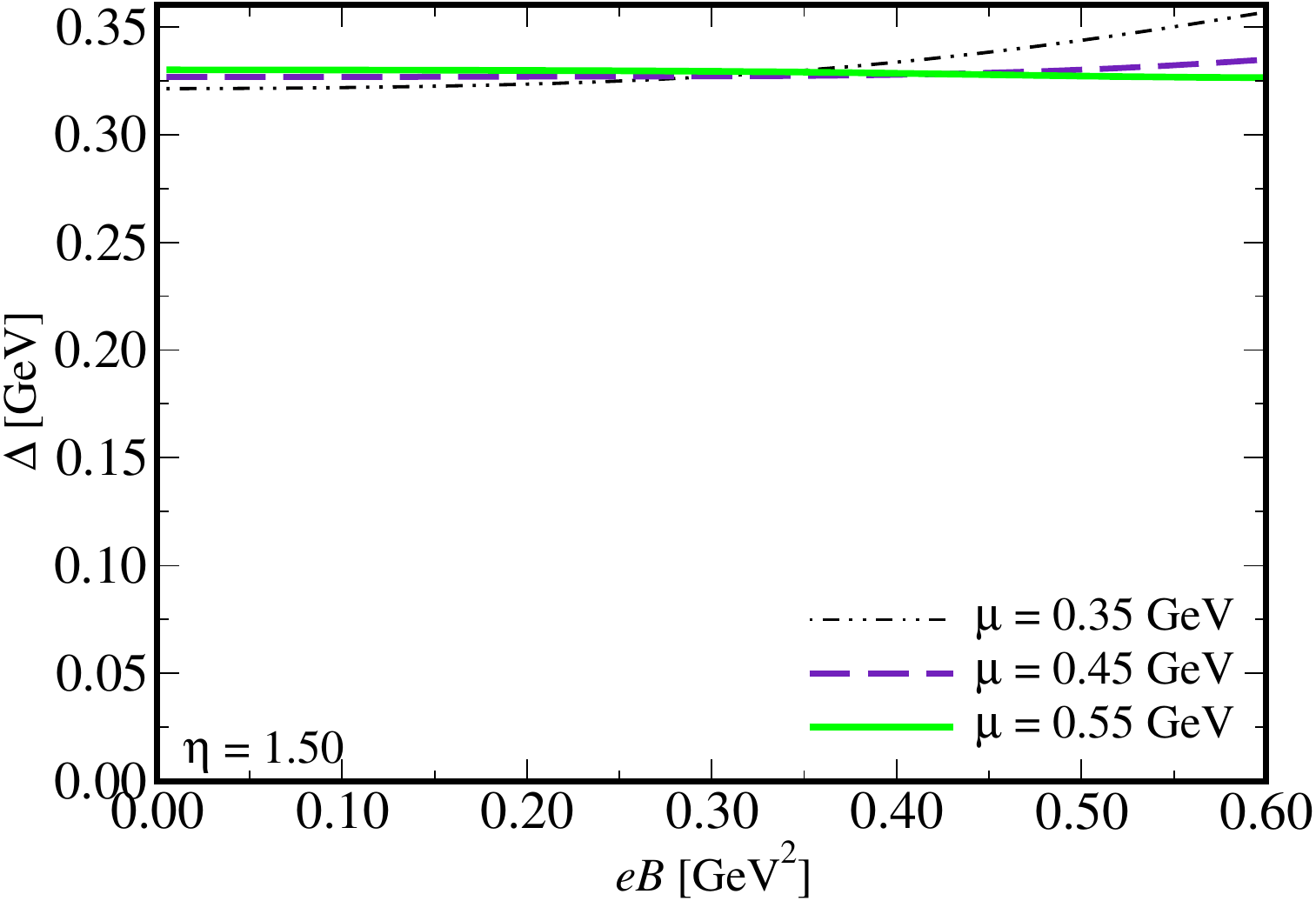}}
  \caption{Constituent quark mass $M$ (\textbf{left} panels) and diquark condensate $\Delta$ (\textbf{right} panels) as a function of the magnetic field $eB$ within the MSS + MFIR method at finite chemical potential. The results are presented for three different values of the ratio between couplings $\eta$.}
   \label{Fig3}
  \end{figure}

Figure~\ref{Fig3} presents the behavior of the constituent quark mass $M$ (left panels) and the diquark condensate $\Delta$ (right panels) as functions of the magnetic field $eB$ within the \mbox{MSS + MFIR} scheme, for different values of the chemical potential $\mu$ and coupling ratio $\eta$.
From the left panels, it is evident that the constituent quark mass $M$ decreases significantly as $\mu$ increases, signaling the partial restoration of chiral symmetry at higher densities. Concurrently, the right panels show that the diquark condensate $\Delta$ is strongly enhanced as $\mu$ increases, indicating a robust color-superconducting phase in the high-density regime. This opposite behavior illustrates the well-known competition between the chiral and diquark condensates. Regarding the magnetic-field dependence, the characteristic de Haas--van Alphen oscillations, which arise as a direct consequence of Landau quantization, are clearly visible in the low-density regime (up to $\mu = 0.45\text{ GeV}$), particularly for smaller values of $\eta$ in the mass plots.
However, as $\mu$ increases to $0.55\text{ GeV}$, these oscillations become strongly suppressed and the curves become remarkably smooth throughout the entire $eB$ range. Physically, this suppression is closely related to the strong growth of the diquark condensate $\Delta$ at larger chemical potentials; the formation of a robust superconducting gap ($\Delta \sim 0.3\text{ GeV}$) stabilizes the Fermi surface and screens thermodynamic quantities from the oscillatory effects associated with the filling of Landau levels, even in the presence of intense magnetic fields. \textcolor{black}{This behavior is consistent with the expected suppression of the Haas–van Alphen oscillations by strong pairing~\cite{Duarte:2015ppa}.}
Furthermore, the coupling ratio $\eta$ plays an important role in determining this balance. As $\eta$ increases from $0.75$ to $1.50$, the values of $\Delta$ increase further for all considered values of $\mu$, driving the system deeper into the superconducting phase.
Consequently, for $\eta = 1.50$, the diquark condensate $\Delta$ becomes much less sensitive to the magnetic field, remaining nearly constant throughout the analyzed range. In this regime, the de Haas–van Alphen oscillations are almost completely washed out.
We restricted our analysis to these values of $\mu$ because higher values would cause the constituent mass $M$ to drop below the current quark mass, signaling an unphysical regime within the present parametrization. Moreover, in the presence of strong magnetic fields, the competing effects of $\mu$ and $eB$ lead to a nontrivial evolution of the constituent mass. These features indicate that the model is being pushed beyond its domain of applicability, preventing a reliable extrapolation to larger chemical potentials.

\begin{figure}[htpb!] 
\centering
{\includegraphics[scale=0.3]{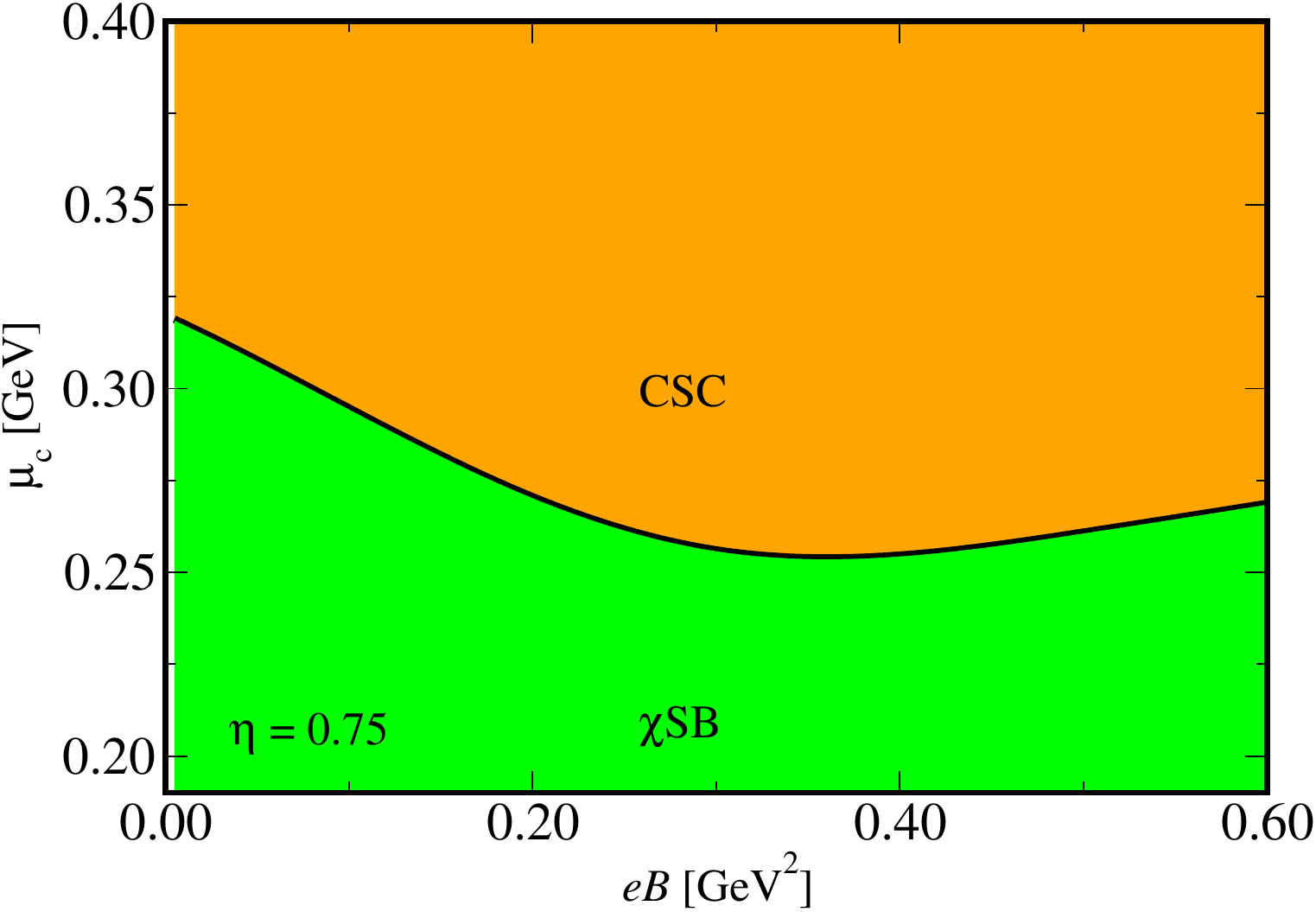}}
{\includegraphics[scale=0.3]{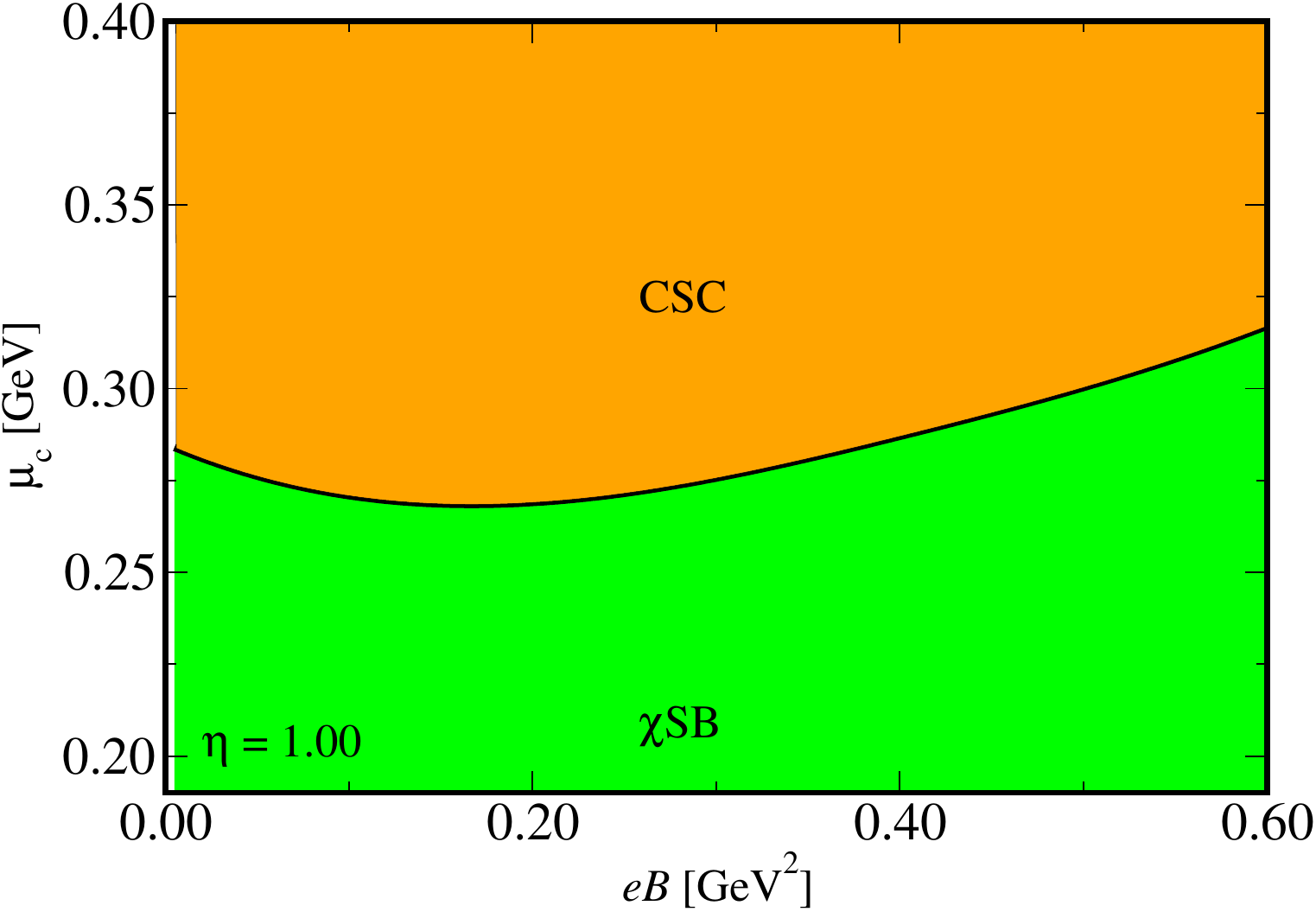}}\vspace{8mm}  
{\includegraphics[scale=0.3]{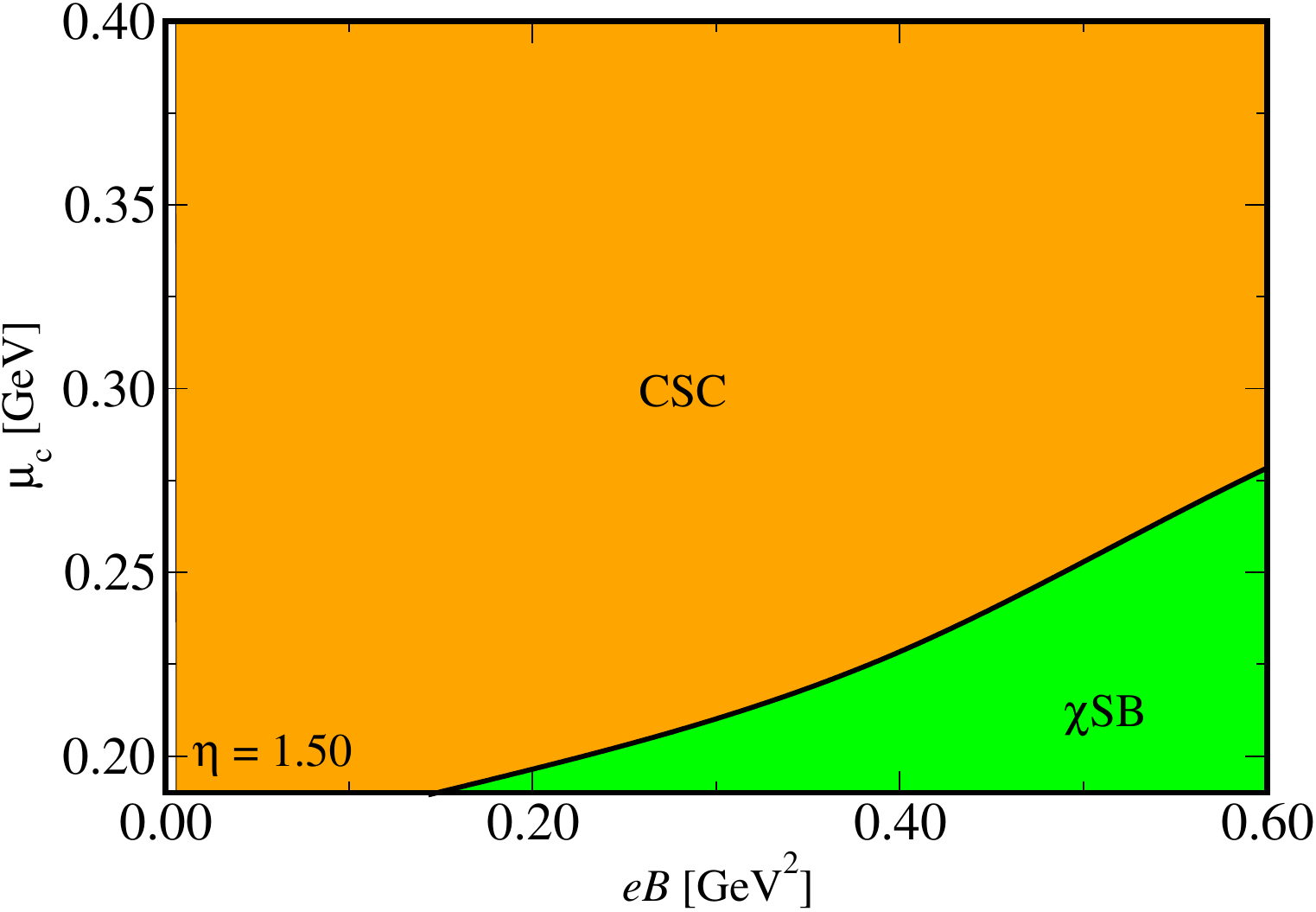}}
  \caption{Phase diagram in the $eB \times \mu$ plane within the MSS + MFIR method. The results are presented for three different values of the coupling ratio $\eta$. These diagrams display the chiral symmetry breaking phase ($\chi$SB) and the color superconducting phase (CSC).}
   \label{Fig4}
  \end{figure}

The phase structure in the $eB \times \mu_c$ plane, obtained within the MSS + MFIR scheme, is presented in Figure~\ref{Fig4}.
The diagram highlights the boundary separating the chiral symmetry breaking ($\chi\text{SB}$) phase from the color superconducting ($\text{CSC}$) phase for three distinct values of the coupling ratio $\eta$, with the transition between these regimes being of first order.
For the lowest coupling ratio ($\eta = 0.75$), the critical chemical potential $\mu_c$ initially decreases as the magnetic field strength increases, reaching a minimum around $eB \approx 0.35\text{ GeV}^2$ before exhibiting a mild turnaround. As $\eta$ increases to $1.00$, this minimum shifts toward lower magnetic fields and the $\text{CSC}$ phase expands.
Finally, for $\eta = 1.50$, the $\text{CSC}$ phase becomes strongly dominant, occupying the low-$\mu_c$ region even at vanishing magnetic fields, while $\mu_c$ exhibits an almost monotonic increase as a function of $eB$ throughout most of the analyzed range. This behavior demonstrates that increasing the coupling ratio significantly favors the onset of color superconductivity over chiral symmetry breaking across the entire magnetic-field range.
This figure also underscores the crucial role played by a proper treatment of divergences when medium-dependent effects are consistently incorporated.
Specifically, regularization schemes that neglect such effects, such as the TRS prescription, artificially predict a normal phase at high chemical potentials in which color superconductivity is entirely absent~\cite{Fayazbakhsh:2010gc}.
Such a result constitutes a well-known regularization artifact that contradicts both established effective-model predictions and the expected physical behavior of cold, dense quark matter. In the high-$\mu$ and low-temperature regime, the system is fundamentally governed by the Cooper instability~\cite{Cooper:1956zz,Collins:1974ky}, which ensures that the formation of diquark Cooper pairs corresponds to the energetically favored state of deconfined quark matter. Consequently, the phase boundaries obtained here reaffirm that the MSS + MFIR framework successfully restores the physically expected phase structure.

\begin{figure}[htpb!] 
{\includegraphics[scale=0.3]{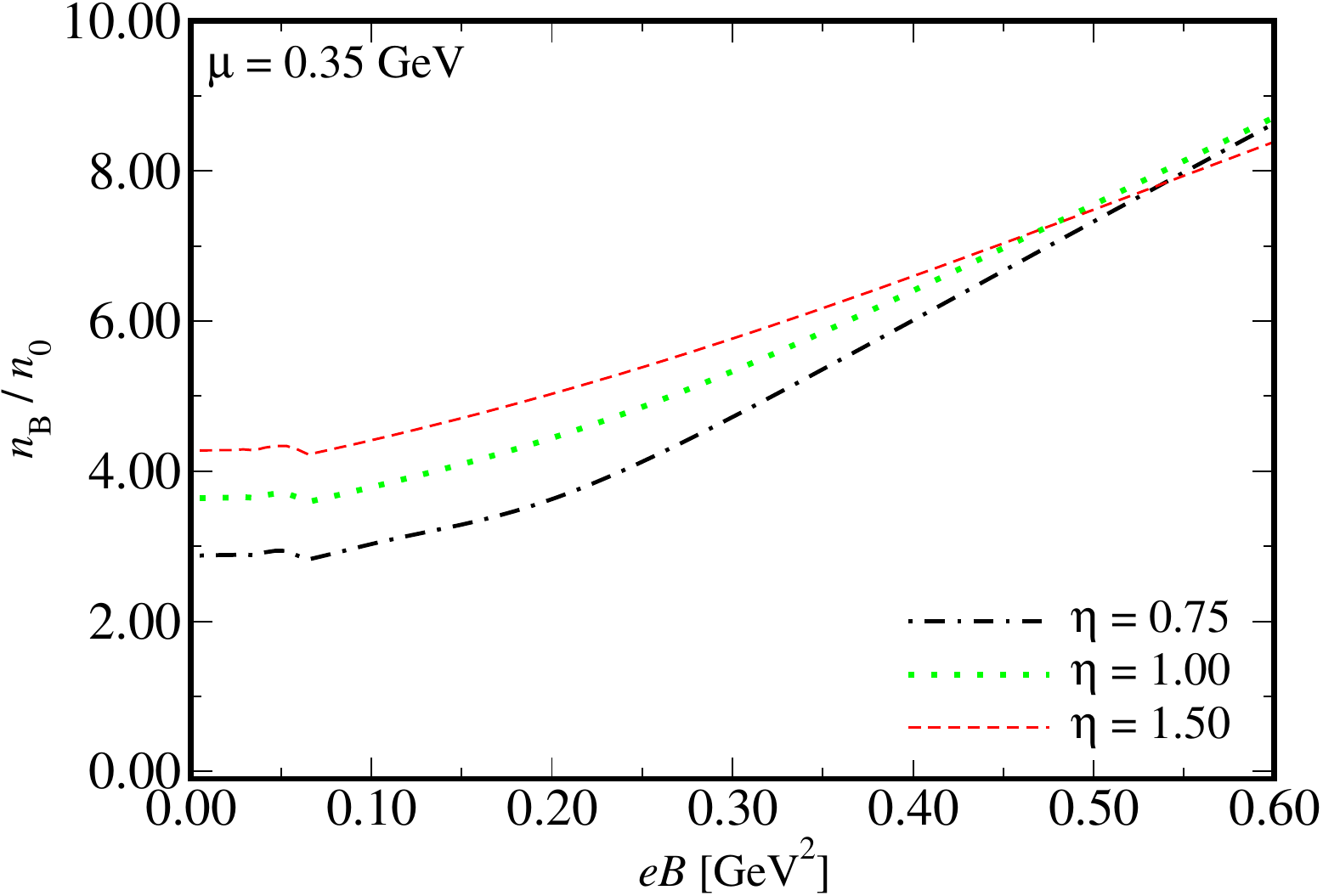}}
{\includegraphics[scale=0.3]{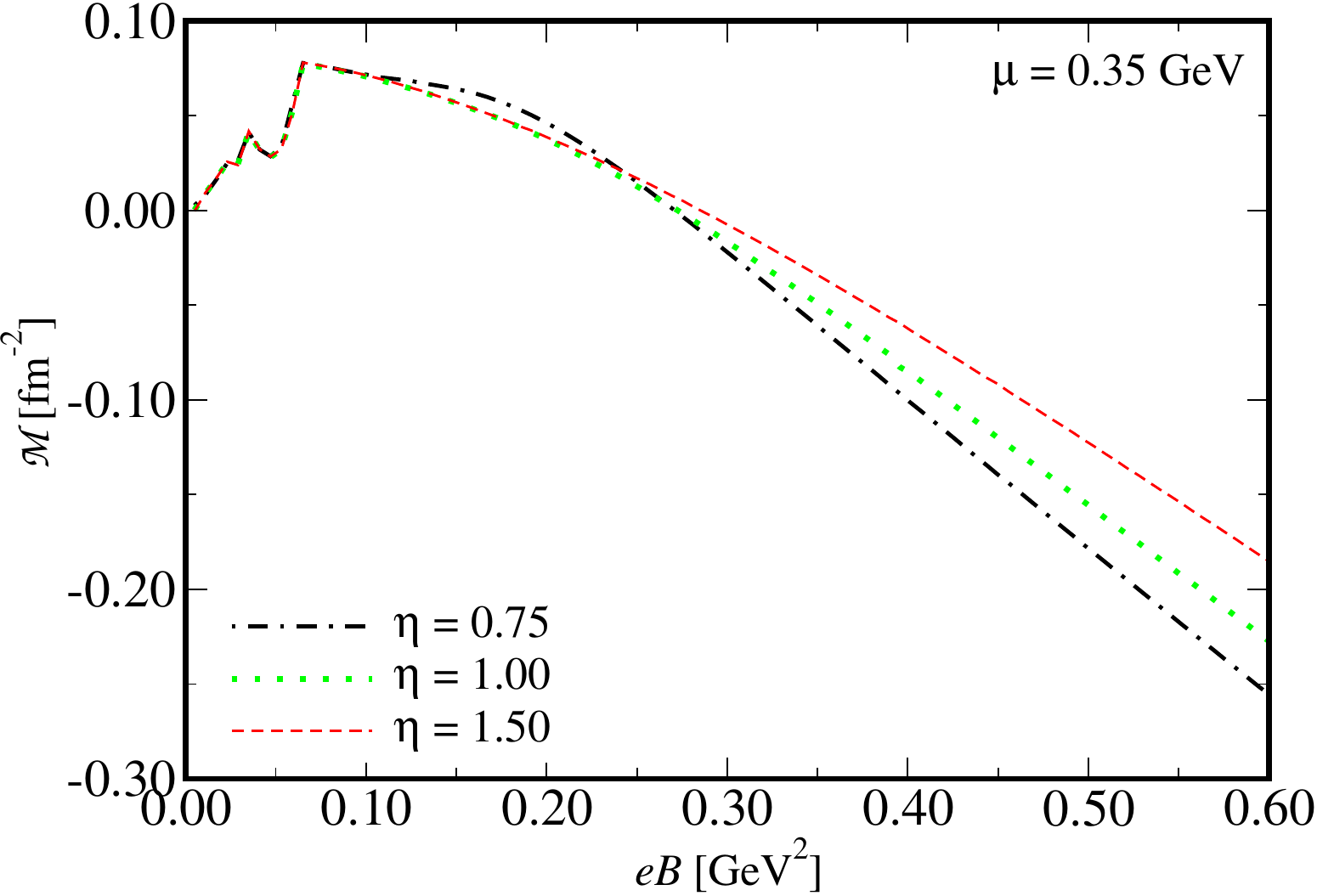}}\\
{\includegraphics[scale=0.3]{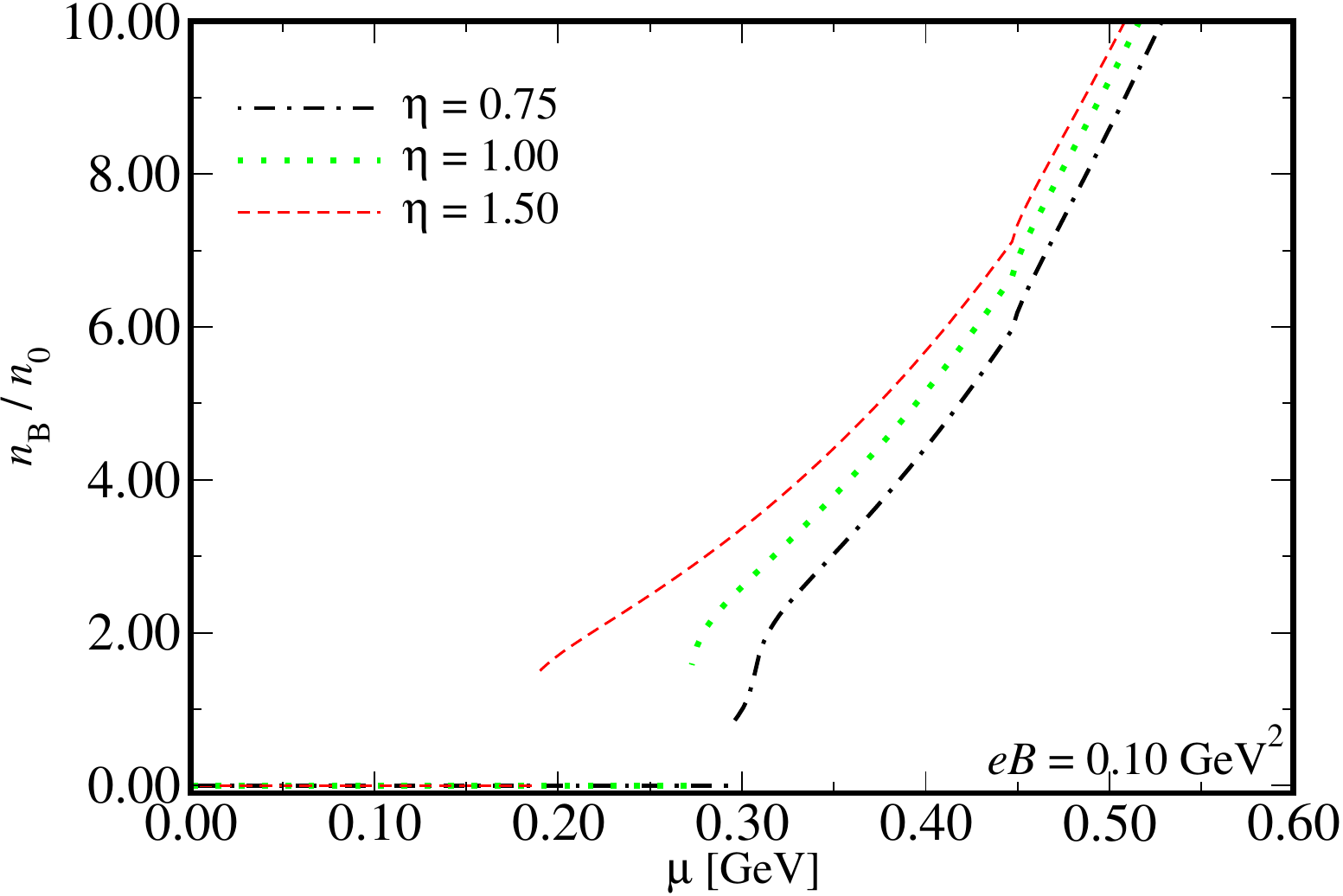}} 
{\includegraphics[scale=0.3]{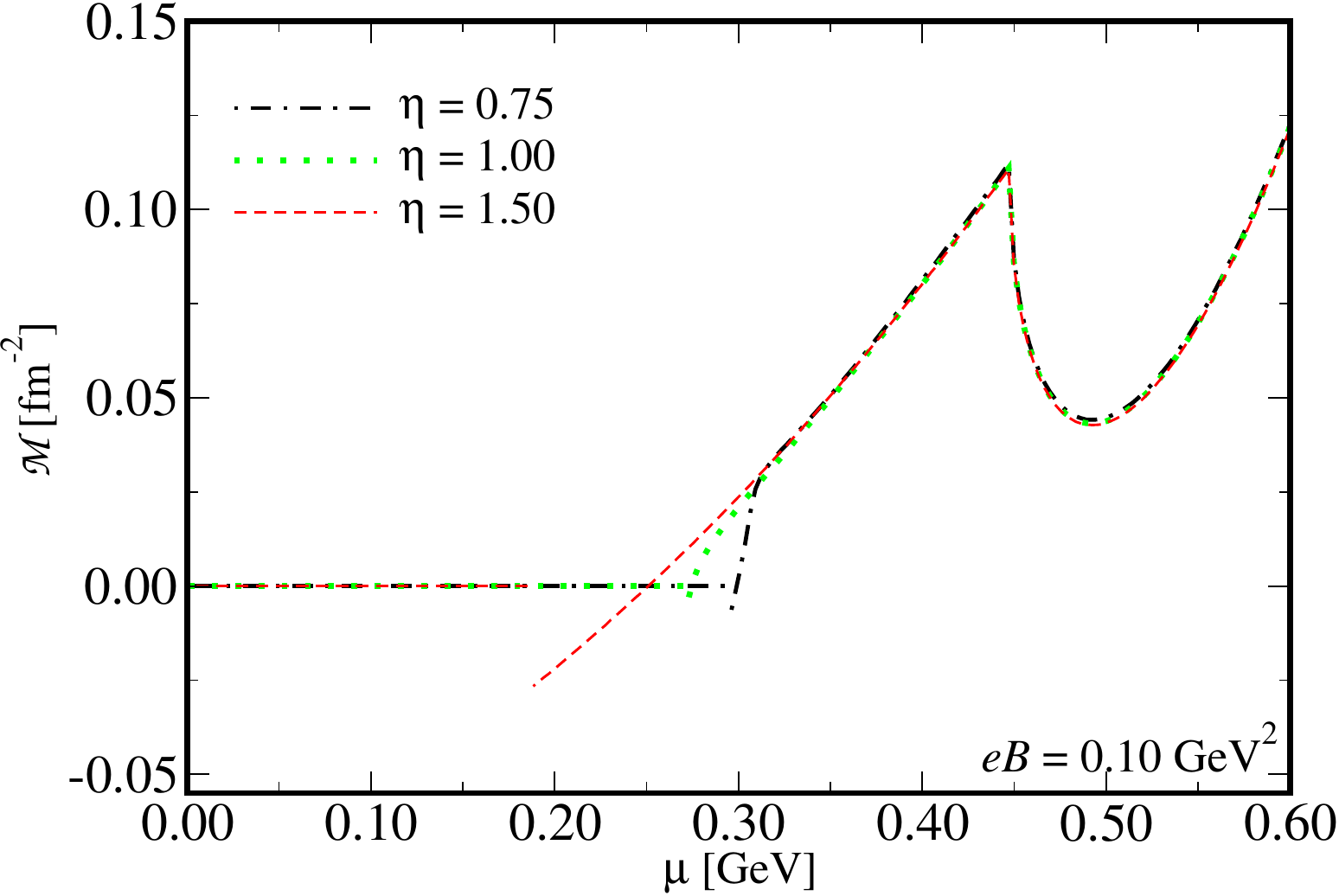}}
  \caption{Normalized baryon number density $n_B/n_0$ (\textbf{left} panels) and magnetization \text{{\termchancery M} } (\textbf{right} panels) as functions of the magnetic field strength $eB$ (\textbf{upper} panels) and the chemical potential $\mu$ (\textbf{lower} panels), calculated within the MSS + MFIR scheme.}
   \label{Fig5}
  \end{figure}

The results shown in Figure~\ref{Fig5} illustrate the thermodynamics of dense quark matter within the MSS + MFIR framework, providing further insight into the interplay between chemical potential, magnetic field, and the coupling ratio $\eta$. As observed in the lower panels, increasing the chemical potential $\mu$ triggers the transition to the CSC phase. The onset of this transition is clearly sensitive to $\eta$, with larger values (e.g., $\eta = 1.50$) shifting the transition toward lower chemical potentials.
Below the critical threshold, both the magnetization \text{{\termchancery M}} and the normalized baryon number density $n_B/n_0$ vanish. Once the dense regime is reached, $n_B/n_0$ increases continuously, while \text{{\termchancery M}} exhibits pronounced de~Haas--van~Alphen oscillations.
The sharp peak followed by a sudden decrease around $\mu \approx 0.45\,\text{GeV}$ is a direct manifestation of Landau quantization, marking the threshold at which a specific Landau level becomes completely occupied.
On the other hand, varying the magnetic-field strength $eB$ (upper panels) induces a nontrivial magnetic response.
For the weaker values of magnetic field used in this plot ($eB < 0.10\,\text{GeV}^2$), the upper-right panel reveals an initial paramagnetic behavior characterized by a pronounced local peak. As the field strength increases, the system undergoes a transition to a strongly diamagnetic regime, where \text{{\termchancery M}} decreases almost linearly at large magnetic fields. Interestingly, the magnetization curves cross near $eB \approx 0.15\,\text{GeV}^2$, beyond which larger values of $\eta$ appear to soften the diamagnetic response. At the same time, the upper-left panel shows that stronger values of magnetic fields reduce the number of occupied Landau levels while increasing their degeneracy, leading to a monotonic enhancement of the baryon number density. While larger values of $\eta$ shift the system toward denser states throughout most of the considered range, a reversal trend occurs in the strong magnetic field regime ($eB \gtrsim 0.50 \text{ GeV}^2$), where the curves cross and lower values of $\eta$ yield higher densities. These findings suggest that the main physical features are robust across different effective descriptions, as the qualitative behavior of the magnetization as a function of chemical potential is consistent with previous results obtained for cold magnetized quark matter within the nonlocal Nambu--Jona-Lasinio model~\cite{Ferraris:2025fva}.

\begin{center}
\begin{figure}[htpb!]
{\includegraphics[scale=0.3]{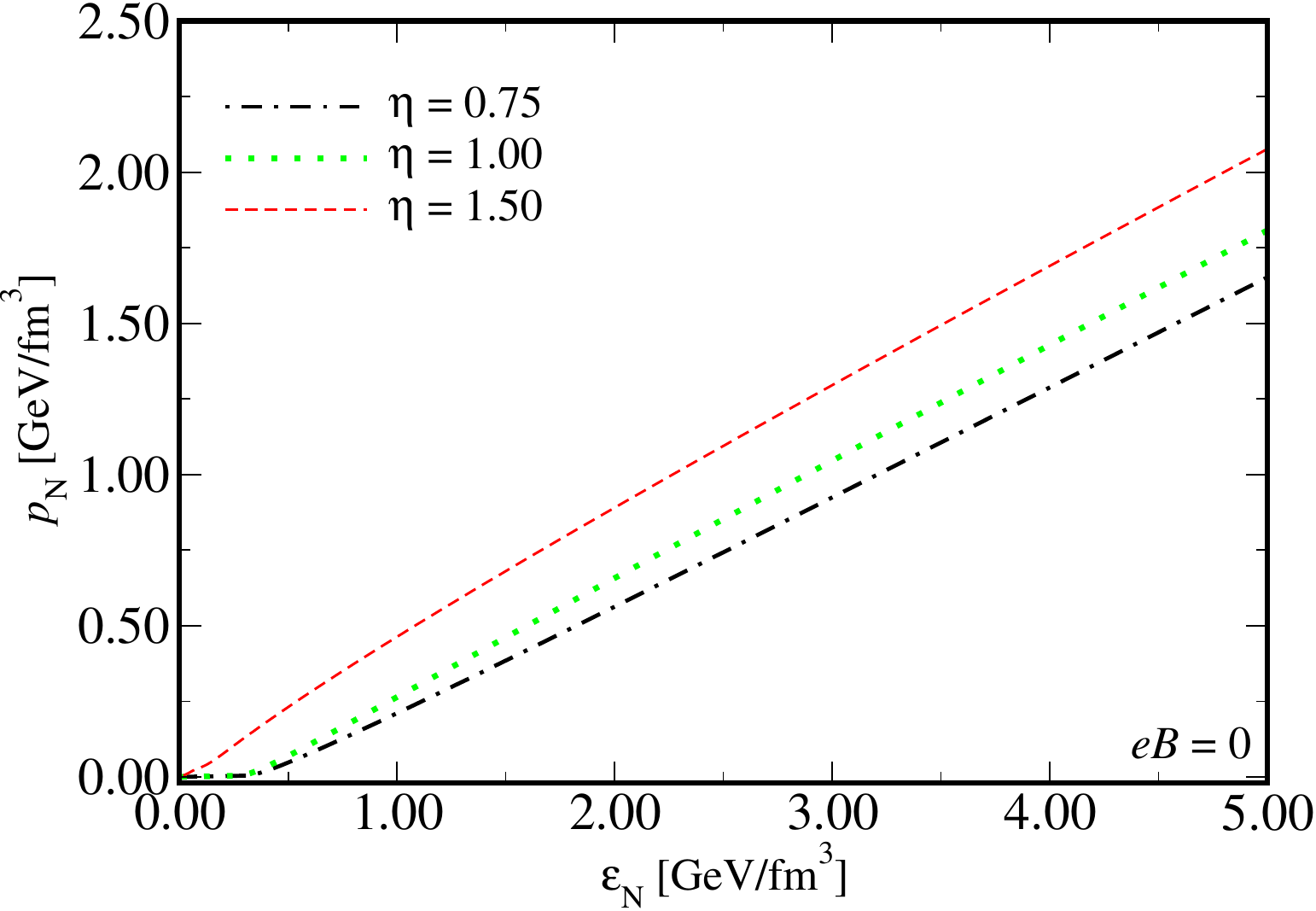}} 
{\includegraphics[scale=0.3]{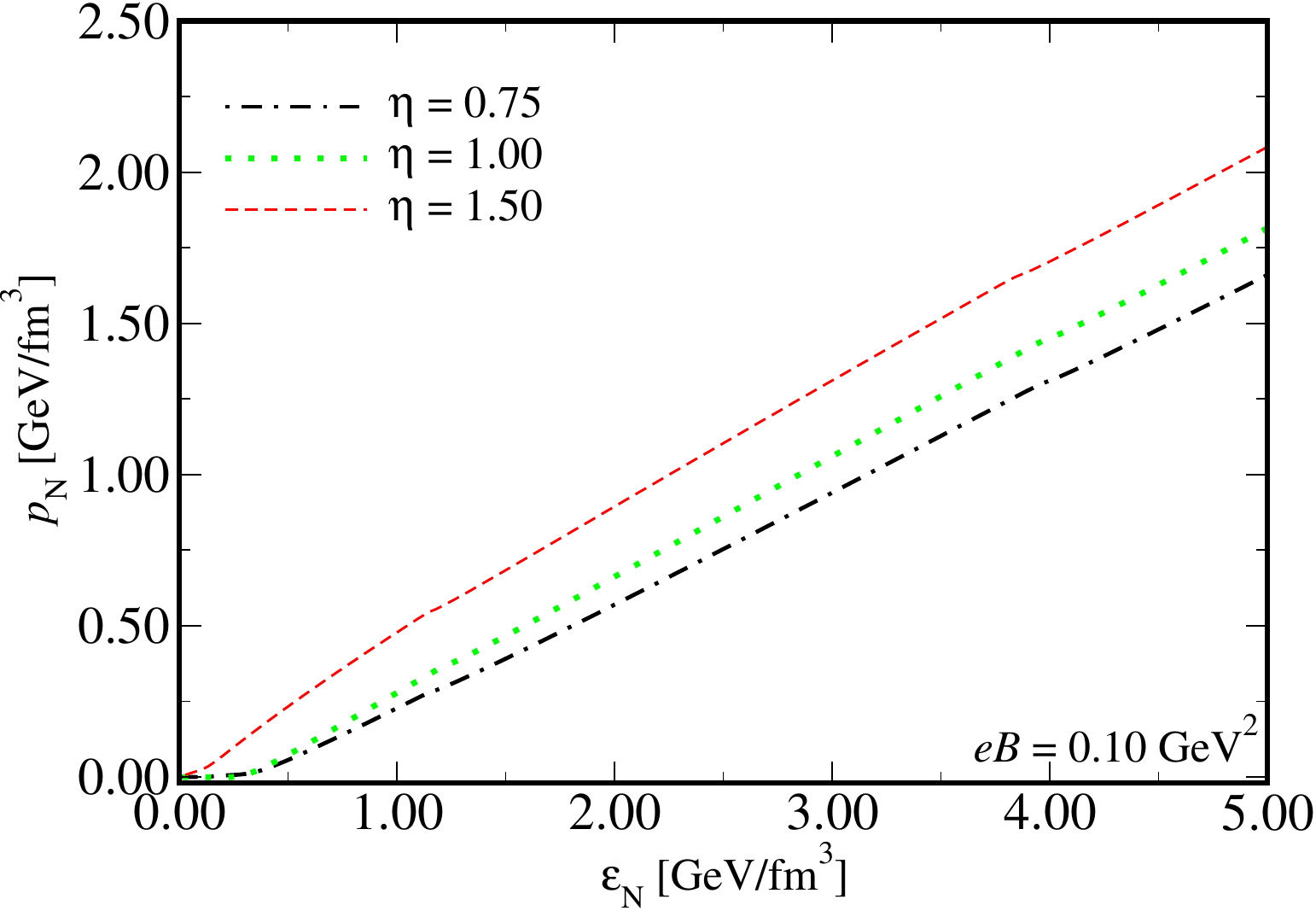}}\\ 
{\includegraphics[scale=0.3]{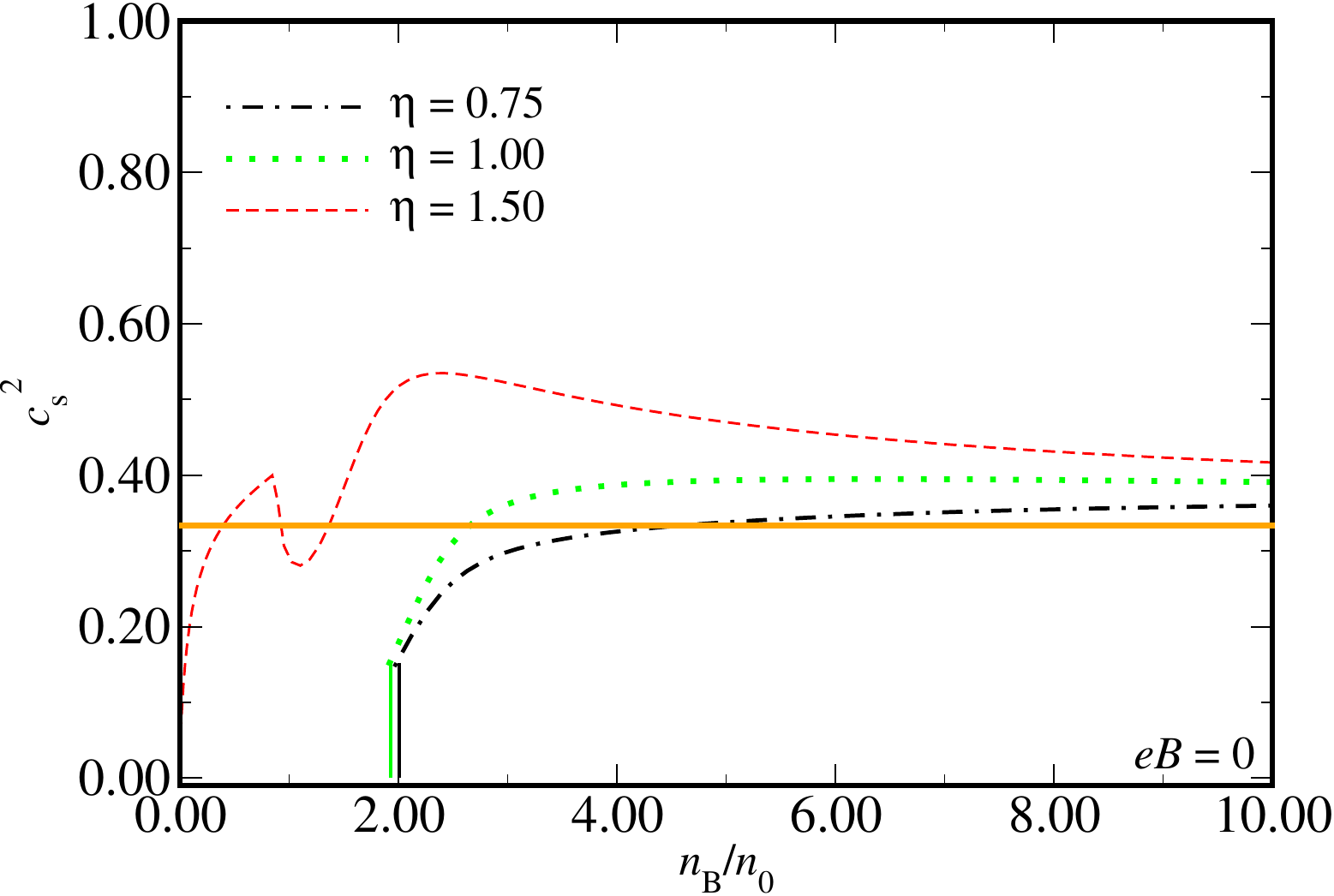}} 
{\includegraphics[scale=0.3]{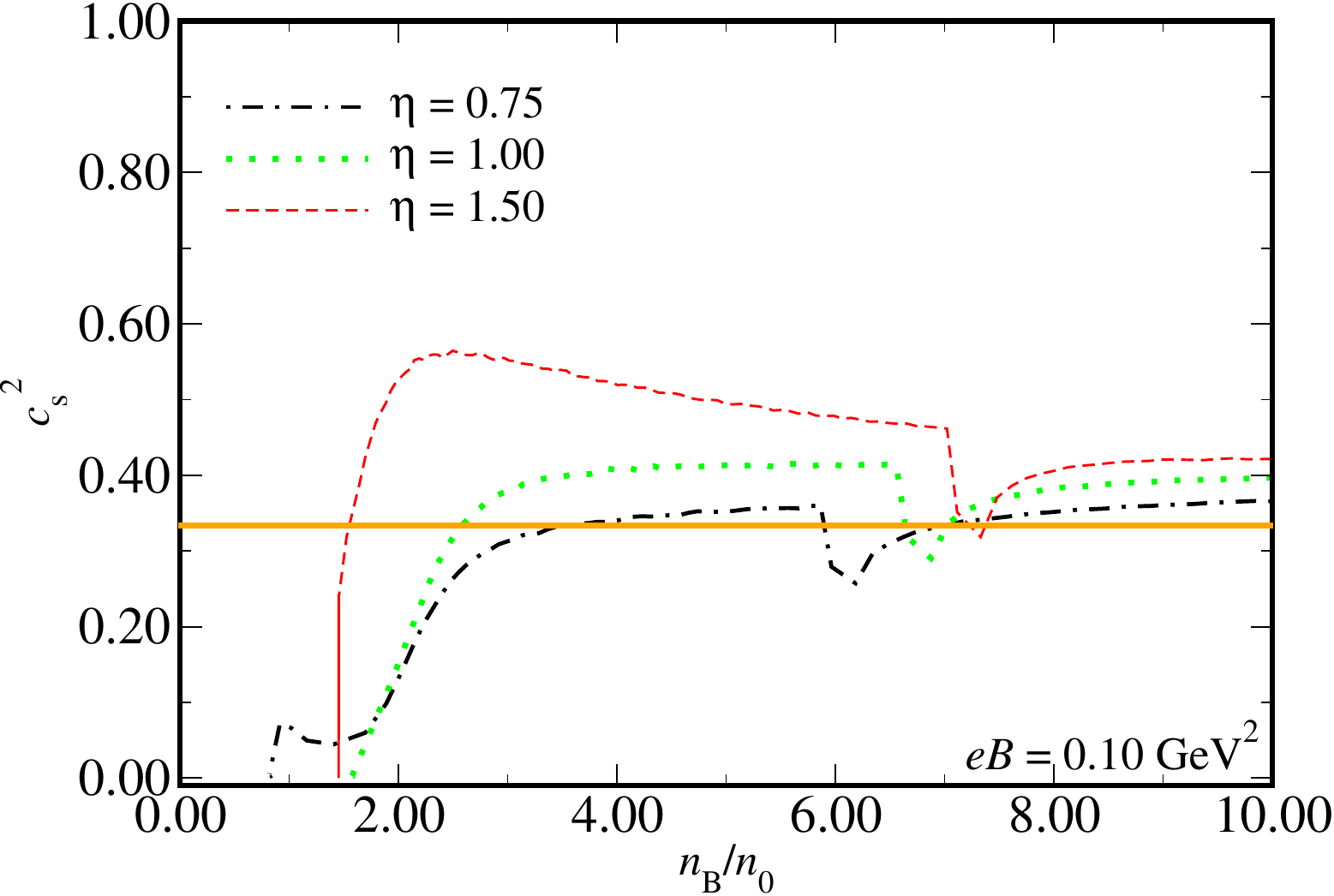}}
  \caption{The Equation of State (EoS)  (\textbf{upper} panels), and the squared speed of sound $c_s^2$ as a function of $n_B/n_0$ (\textbf{lower} panels). The \textbf{left} panels show results within the MSS without MFIR. The \textbf{right} panels are in MSS + MFIR scheme. }
   \label{Fig6}
  \end{figure}
 \end{center} 

Figure~\ref{Fig6} presents the Equation of State (EoS), together with the squared speed of sound $c_s^2$, for both vanishing (left panels) and finite magnetic fields (right panels). The upper panels indicate that the EoS becomes progressively stiffer as the parameter $\eta$ increases, leading to higher pressures at a fixed energy density.
Regarding the speed of sound (lower panels), $c_s^2$ exhibits a nonmonotonic behavior, with a pronounced peak that exceeds the conformal limit of $1/3$ for larger values of $\eta$. In the absence of a magnetic field (bottom-left panel), where $c_s^2$ is shown as a function of the normalized baryon density $n_B/n_0$, the system undergoes a second-order phase transition for $\eta = 1.50$.
In contrast, for the magnetized case (bottom-right panel), the transition at $\eta = 1.50$ becomes clearly first-order, as evidenced by the sharp discontinuity in the speed of sound. Finally, it is worth noting that for $\eta = 0.75$ and $\eta = 1.00$, the transitions are of first order in both scenarios; however, in the finite-magnetic-field case (bottom-right panel), the corresponding discontinuities are strongly suppressed and cannot be visually resolved within the current scale of the plots.

It is worth emphasizing that the violation of the conformal value observed at intermediate densities does not imply that $c_s^2$ should remain above $1/3$ asymptotically. Within the MSS, finite medium contributions are left unregularized, and previous applications show that the peak is followed by a tendency toward the conformal value at increasing density~\cite{XavierdeAzeredo:2026wlq}. At fixed magnetic field, magnetic corrections are also expected to become subleading for $\mu^2 \gg eB$. Thus, the expected high-density trend is $c_s^2 \rightarrow 1/3$, typically from above in effective-model calculations~\cite{Azeredo:2026wgl,Andersen:2026sjc,Gholami:2024ety}. Nevertheless, since the NJL model is nonrenormalizable and does not incorporate QCD asymptotic freedom, this limiting behavior should be understood as a qualitative trend rather than as a controlled prediction in the strict $\mu \rightarrow \infty$ limit.

\begin{center}
\begin{figure}[htpb!] 
{\includegraphics[scale=0.3]{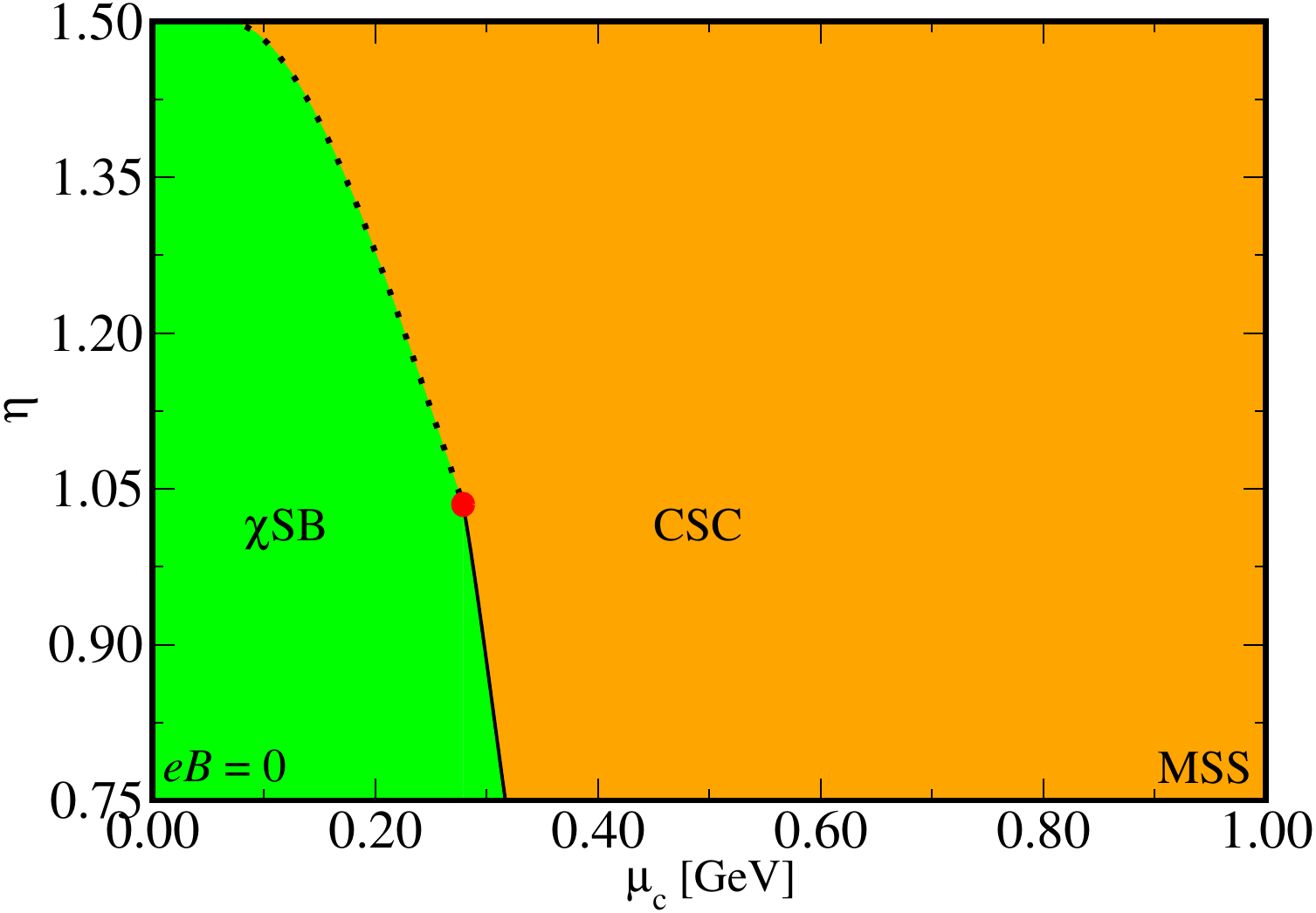}} 
{\includegraphics[scale=0.3]{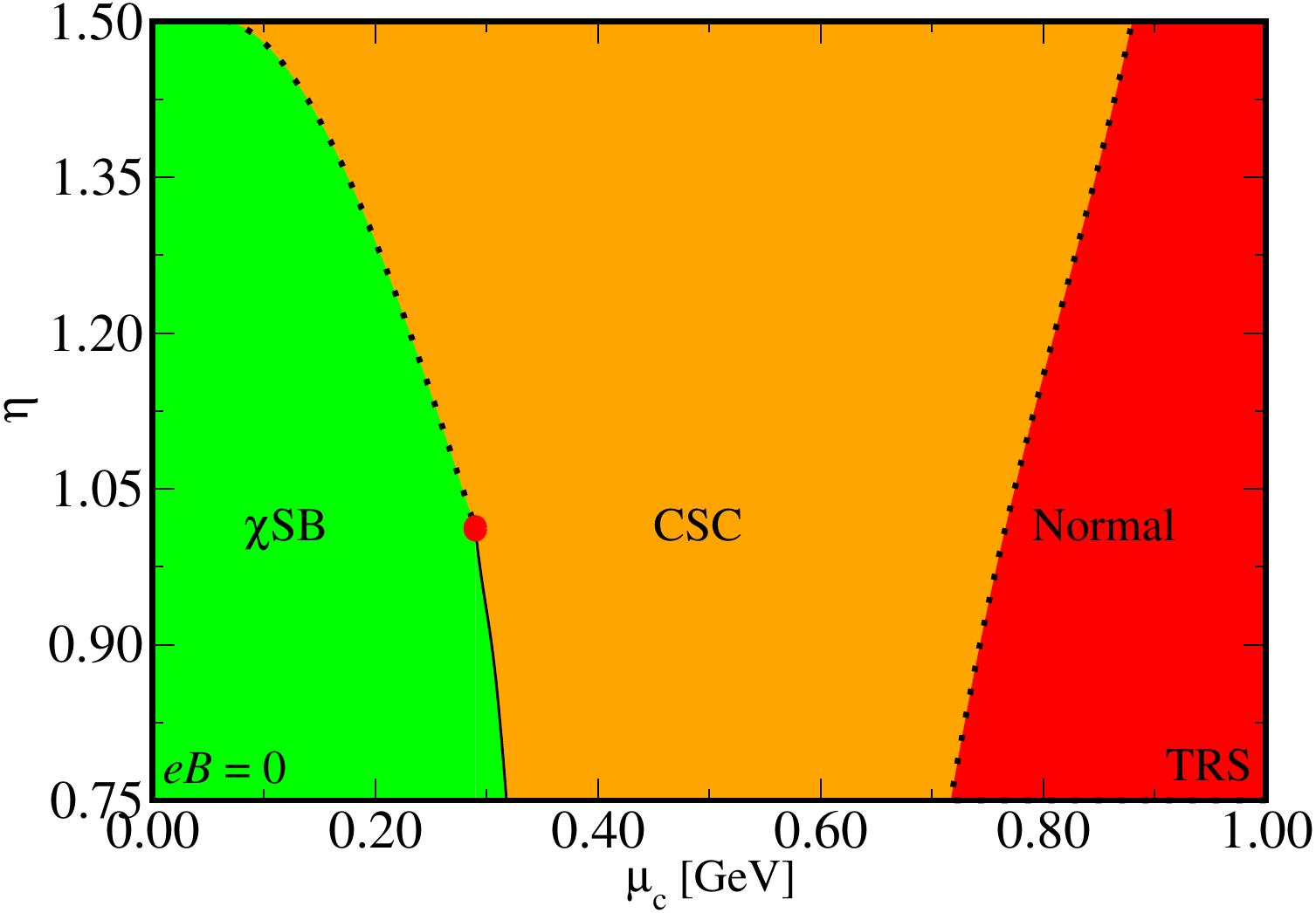}} 
\\
{\includegraphics[scale=0.3]{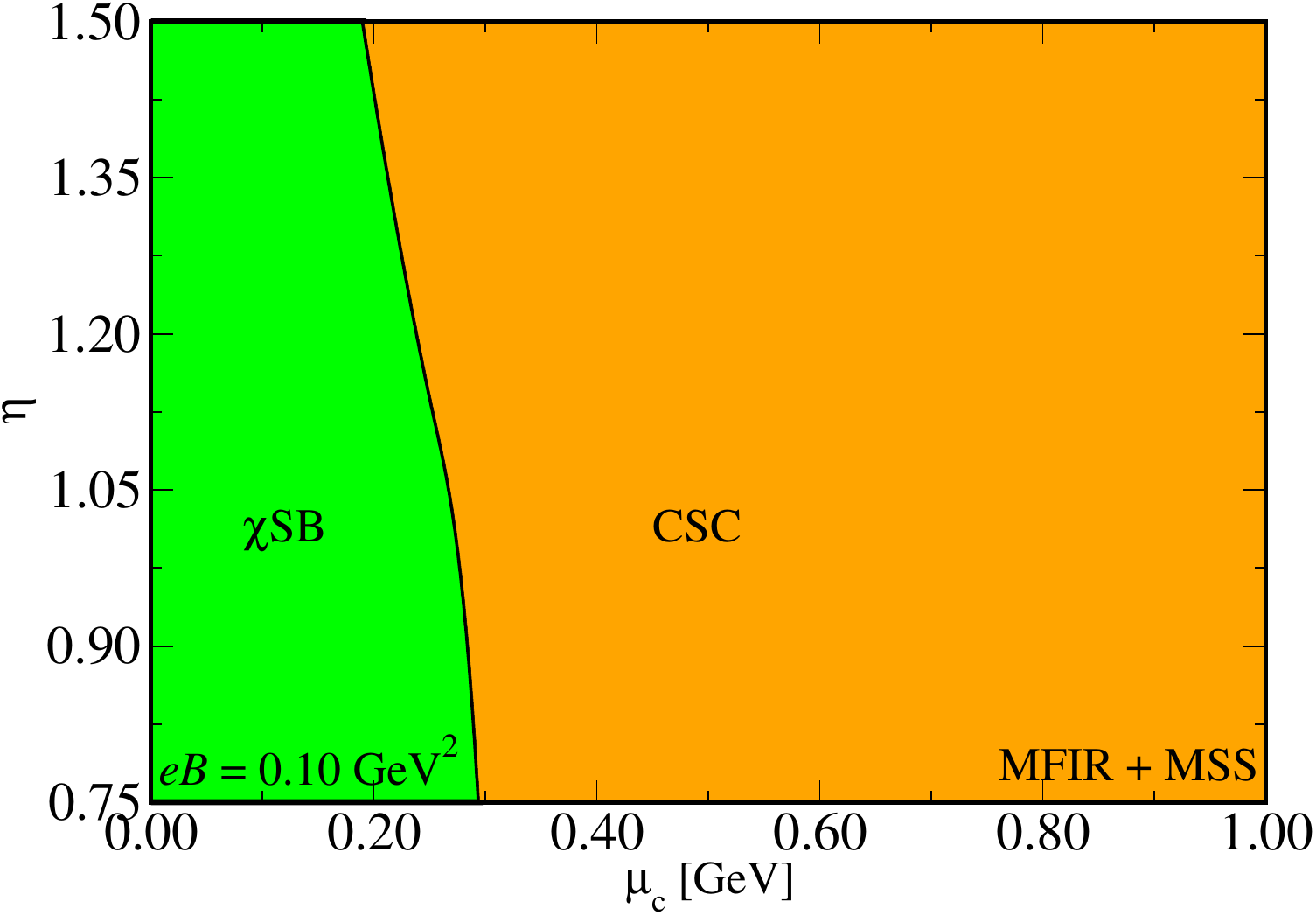}} 
{\includegraphics[scale=0.3]{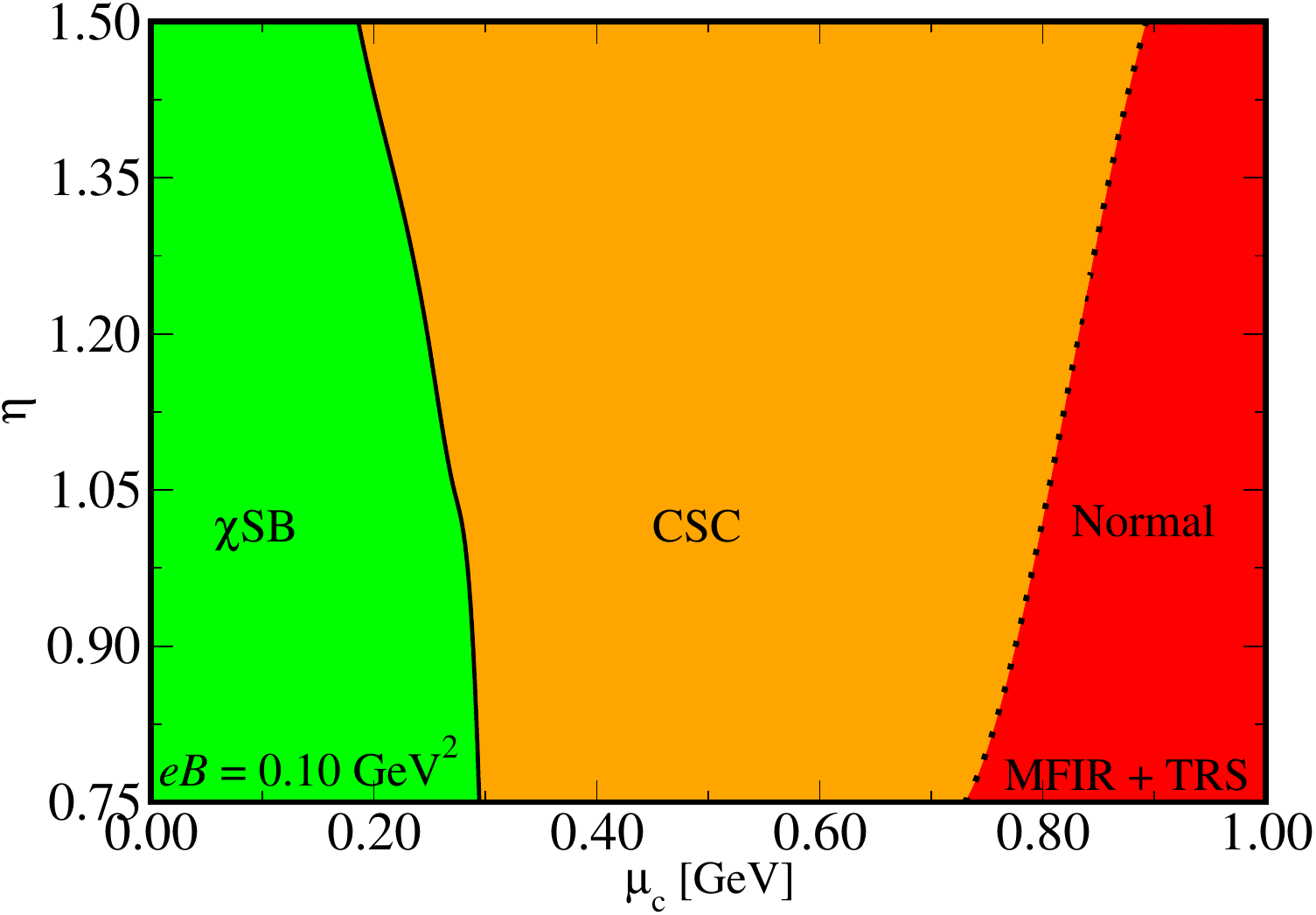}}
  \caption{Phase diagram in the $\eta \times \mu$ plane, comparing the MSS and TRS schemes at zero magnetic field (\textbf{top} panels) and at a finite magnetic field using the MSS + MFIR and TRS + MFIR methods (\textbf{bottom} panels). The diagrams display the chiral symmetry breaking ($\chi\text{SB}$), color superconducting phase (CSC), and a phase with restored chiral symmetry (normal). The red dot indicates the \textcolor{black}{tricritical point (TP)}.}
   \label{Fig7}
  \end{figure}
 \end{center} 

Finally, Figure~\ref{Fig7} displays the phase diagrams in the $\eta \times \mu$ plane, illustrating the impact of vanishing (top panels) and finite magnetic fields (bottom panels) within the MSS and TRS schemes.
For the zero-magnetic-field case ($eB = 0$), both the MSS and TRS frameworks exhibit a transition from the chiral-symmetry-breaking $\chi\text{SB}$ phase to the color-superconducting CSC phase. This characteristic competition and subsequent transition between chiral restoration and diquark condensation at high densities are consistent with the fundamental phase-diagram structure established in early effective-model studies~\cite{Berges:1998rc}.
For larger values of the coupling ratio $\eta$, this transition is of second order (represented by the dotted lines) and terminates at a Tricritical Point (TP), indicated by the red dots, below which the transition becomes first order (solid lines). A major qualitative distinction between the two regularization schemes emerges at high chemical potentials. Within the MSS scheme, the CSC phase remains stable throughout the entire high-density region, a behavior that is in close agreement with perturbative QCD and renormalization group calculations~\cite{Son:1998uk,Gholami:2024diy,Gholami:2024ety,Gholami:2025guq}. Conversely, the TRS scheme fails to sustain this phase at very high densities; the diquark condensate vanishes as $\mu$ increases, forcing the system to return to the chirally restored (normal) phase, with this transition also being of second order. When a finite magnetic field is introduced ($eB = 0.10\text{ GeV}^2$, bottom panels) through the MFIR framework, significant modifications emerge. The most prominent feature in both the \text{MSS + MFIR} and \text{TRS + MFIR} approaches is that the transition from the $\chi\text{SB}$ phase to the CSC phase becomes entirely first order throughout the whole range of $\eta$ considered. Consequently, the second-order transition lines and their corresponding TPs disappear under the influence of the magnetic field, while the characteristic high-$\mu$ asymptotic behavior of each regularization scheme (MSS versus TRS) remains qualitatively unchanged.

The disappearance of the second-order segment at finite magnetic field deserves some qualification. Functional-renormalization-group studies of the chiral transition at finite $T$, $\mu$, and $B$ have found that increasing the magnetic field can move the \textcolor{black}{tricritical point} toward higher temperatures and lower chemical potentials, effectively enlarging the first-order region of the chiral phase boundary~\cite{Andersen:2012bq,Ruggieri:2014bqa}. This provides qualitative support for the tendency observed here, although such calculations do not include a color-superconducting order parameter and therefore do not directly describe the $\chi$SB→CSC transition. Likewise, functional studies of color superconductivity indicate that fluctuations can favor a first-order superconducting transition, but this result has not yet been established in the presence of an external magnetic field. A direct comparison with lattice QCD is presently not possible, since the cold, baryon-dense regime relevant for color superconductivity remains inaccessible in physical three-color QCD because of the sign problem. We also emphasize that a magnetic-field-induced strengthening of the first-order transition is not universal among effective models: the detailed phase structure can depend strongly on neutrality constraints, the diquark coupling, Landau-level dynamics, and the regularization prescription. In the present calculation, however, the fact that the second-order segment disappears in both the MSS + MFIR and TRS + MFIR schemes indicates that this qualitative change is associated primarily with the magnetic field, rather than with the treatment of the medium-dependent contributions.

\section{Summary and Outlook}
\label{sec:discussions}
In this review, we revisited the role of regularization procedures in the description of cold and dense magnetized quark matter within effective models of QCD, with particular emphasis on the Magnetic-Field-Independent Regularization (MFIR) and the Medium Separation Scheme (MSS).
We discussed how traditional regularization procedures may artificially entangle vacuum, magnetic-field and medium contributions, leading to unphysical effects such as spurious oscillations and the artificial suppression of superconducting gaps at high chemical potentials. In contrast, the MFIR and MSS prescriptions consistently isolate the ultraviolet-divergent vacuum sector from finite magnetic and medium contributions, leading to a more physically reliable thermodynamic description.
Within this framework, we analyzed the behavior of the constituent quark mass, superconducting gap, magnetization, baryon density, phase structure, equation of state, and speed of sound under strong magnetic fields and finite density. In particular, we verified that the superconducting gap remains finite at high densities, even in the presence of strong magnetic fields, eliminating the appearance of an unphysical normal phase commonly found in traditional regularization schemes.
The results reviewed here reinforce the importance of properly disentangling vacuum and medium effects in nonrenormalizable effective models of QCD. 
This separation appears to be essential for preserving the physically expected behavior of dense strongly interacting matter, especially in regimes where medium effects dominate the thermodynamics, while simultaneously avoiding artifacts associated with an incomplete treatment of vacuum and medium contributions.
Finally, the good qualitative agreement of MSS-based results with renormalizable approaches, RG-consistent frameworks, and lattice-inspired results in several QCD-like systems suggests that these regularization procedures may provide an important guideline for the construction of more reliable effective descriptions of dense strongly interacting matter under extreme conditions relevant to compact stars, heavy-ion collisions, and QCD-like theories.
More broadly, the studies reviewed here indicate that the separation of vacuum, magnetic and medium contributions is not merely a technical aspect of the regularization procedure, but may qualitatively affect the physical predictions of effective QCD models. In particular, the persistence of color-superconducting phases at high densities, the absence of spurious instabilities, and the improved agreement with renormalizable and lattice-inspired approaches suggest that medium-separated schemes provide a more faithful description of strongly interacting matter under extreme conditions.
Future investigations should extend these ideas to more realistic scenarios, including three-flavor matter, charge-neutral and beta-equilibrated systems, spin-polarized quark matter, inhomogeneous phases, and the simultaneous presence of strong magnetic fields and finite temperature. Such developments may help clarify the phase structure of dense QCD and provide more robust effective descriptions of matter in compact stars and other extreme astrophysical and laboratory environments.

\section{Acknowledgements}

This work was partially supported by Conselho Nacional de Desenvolvimento Cient\'ifico e Tecno\-l\'o\-gico (CNPq), Grant Nos. 312032/2023-4, 402963/2024-5, 45182/2024-5 (R.L.S.F.) and 404964/2025-7 (D.C.D and R.L.S.F), as well as Grant Nos. 141270/2023-3 and 201300/2025-7 (B.S.L.), and 200037/2026-9 (W.R.T.); Fundação Carlos Chagas Filho de Amparo à Pesquisa do Estado do Rio de Janeiro (FAPERJ), Grant No.SEI-260003/019544/2022 (W.R.T);
Funda\c{c}\~ao de Amparo \`a Pesquisa do Estado do Rio 
Grande do Sul (FAPERGS), Grants Nos. 24/2551-0001285-0 (R.L.S.F.) and 23/2551-0001591-9 (D.C.D. and J.A.R.S.P.); Coordena\c{c}\~ao de
Aperfei\c{c}oamento de Pessoal de N\'ivel Superior---Brasil (CAPES)---Finance Code 001 (F.X.A.). This work is also part of the project Instituto
Nacional de Ciência e Tecnologia—Física Nuclear e
Aplicações (INCT—FNA), Grants No. 464898/2014-5 and No. 408419/2024-5
and supported by the Serrapilheira Institute (Grant
No. Serra-2211-42230).


\appendix

\section[\appendixname~\thesection]{Analytical and numerical implementation of MFIR on $\Omega_{\frac{1}{2}}$} 
\label{appMFIR}

In this appendix, we show the main steps to obtain the MFIR expression for $\Omega_{\frac{1}{2}}$. Starting from the $T = 0$ limit expression given in the last line of Equation~\eqref{Omega_a},
\begin{equation}
\Omega_{\frac{1}{2}} = \frac{eB}{2\pi^{2}}\sum_{n=0}^{\infty}\alpha_{n}\int\limits_{0}^{\infty}dp_{z} \left(E_{p,\frac{1}{2}}^{+} + E_{p,\frac{1}{2}}^{-} \right) ,  
\end{equation}
one may write
\begin{equation}
\Omega_{\frac{1}{2}} = \frac{eB}{2\pi^{2}}\sum_{n=0}^{\infty}\alpha_{n}\int\limits_{0}^{\infty}dp_{z} F(p_z^2 + neB) 
+ \frac{eB}{\pi^{2}}\sum_{n=0}^{\infty}\alpha_{n}\int\limits_{0}^{\infty}dp_{z} \sqrt{p_z^2 + neB + M^2 + \Delta^2} \, ,
\label{Omega05app}
\end{equation}
where $F(z)$ is defined in Equation~\eqref{Fz}. The first integral above is manipulated by adding and subtracting the $eB = 0$ contribution, and by using the definition of the degeneracy factor, it can be written as
\begin{align}
 \frac{eB}{2\pi^{2}}\sum_{n=0}^{\infty}\alpha_{n}&\int\limits_{0}^{\infty}dp_{z} F(p_z^2 + neB) = 4\int\frac{d^{3}p}{\left(2\pi\right)^{3}}F(\vec{p}^{2})\nonumber\\
&+\frac{2eB}{4\pi^{2}}\int_{-\infty}^{+\infty}dp_{z}\left\{ \frac{F\left(p_{z}^{2}\right)}{2}+\left[\sum_{n=1}^{\infty}F\left(p_{z}^{2}+neB\right)\right]-\int_{0}^{\infty}dy~F\left(p_{z}^{2}+eBy\right)\right\},
\label{intT1}
\end{align}
where the last integral is obtained by expressing the subtracted $eB = 0$ contribution in cylindrical coordinates, introducing the variable $y=\frac{p_{\rho}^{2}}{eB}$, and carrying out the radial and angular integrations. For the second integral in~\eqref{Omega05app}, we follow the same steps addressed in Ref.~\cite{Menezes:2008qt}, by using dimensional regularization and the replacement $M^2\to M^2 + \Delta^2$ to~obtain
\begin{align}
  \frac{eB}{\pi^{2}}\sum_{n=0}^{\infty}\alpha_{n}&\int\limits_{0}^{\infty}dp_{z} \sqrt{p_z^2 + neB + M^2 + \Delta^2}   =8\int\frac{d^{3}p}{(2\pi)^{3}}\sqrt{\vec{p}^{2}+M^{2} + \Delta^2}\nonumber\\
&+\frac{(eB)^{2}}{2\pi^{2}}\left\{ +\zeta^{\prime}\left(-1,x\right)+\frac{(x-x^{2})}{2}\ln\left(x\right)+\frac{x^{2}}{4}\right\},
\label{intT2}
\end{align}
with $x = (M^2 + \Delta^2)/(eB)$.  Collecting~\eqref{intT1} and~\eqref{intT2} in~\eqref{Omega05app} we obtain
\begin{align}
\Omega_{\frac{1}{2}} =& 4\int_{\Lambda} \frac{d^{3}\vec{p}}{(2\pi)^{3}} \left[ \sqrt{\left( \sqrt{\vec{p}^{2} + M^{2}} + \mu \right)^{2} + \Delta^{2}}  \right.  \left. + \sqrt{\left( \sqrt{\vec{p}^{2} + M^{2}} - \mu \right)^{2} + \Delta^{2}}   -2 \sqrt{\vec{p}^{2}+M_0^{2}}\right]\nonumber \\
 & + \frac{eB}{4\pi^{2}} \int_{-\infty}^{+\infty} dp_{z} F\left(p_{z}^{2}\right) + \frac{(eB)^{2}}{2\pi^{2}} \left[ \zeta^{\prime}\left(-1, x\right) + \frac{(x - x^{2})}{2} \ln(x) + \frac{x^{2}}{4} \right] \nonumber \\
 & + \frac{eB}{2\pi^{2}} \int_{-\infty}^{+\infty} dp_{z} \left[\sum_{n=1}^{\infty} F\left(p_{z}^{2} + neB\right)
\right. \left. - \int_{0}^{\infty} dy~ F\left(p_{z}^{2} + eBy\right) \right], 
\end{align}
which is the form shown in Equation~\eqref{W05}.
 
In order to handle the computationally challenging infinite sum in the last line of this equation, we make use of an asymptotic expansion method~\cite{Allen:2015paa}. We begin by defining the magnetic field-dependent contribution $\Omega_{W}$ through the balance between the discrete Landau level sum ($T_S$) and the continuous integral ($T_I$):

\begin{equation}
\Omega_{W} = \frac{eB}{2\pi^{2}}\int_{-\infty}^{+\infty}dp_{z}\left[T_{S}-T_{I}\right] \, ,
\end{equation}
where
\begin{align}
T_{S} &= \sum_{n=1}^{\infty}F\left(p_{z}^{2}+neB\right) \, ,\\
T_{I} &= \int_{0}^{\infty}dy~F\left(eBy+p_{z}^{2}\right) \, .
\end{align}
To evaluate these expressions, we adopt an asymptotic expansion for $F(z)$ at large $z$:
\begin{equation}
F\left(z\right) = \sum_{i=1}^{\infty}\frac{c_{i}}{z^{i+\frac{1}{2}}} \, ,
\label{SeriesF}
\end{equation}
which can be truncated in the first leading terms. In this work we have considered the four first terms for both thermodynamic potential and gap equations.
To implement this approximation safely, we introduce a matching cutoff parameter $m_p$. For values below $m_p$, the exact function $F(z)$ is retained, while the tail distributions are replaced by the power series given in Equation~\ref{SeriesF}.
For the integral contribution $T_I$, splitting the domain yields:
\begin{equation}
T_{I} = \int_{0}^{m_{p}}dy~F\left(eBy+p_{z}^{2}\right) + \sum_{i} c_i \int_{m_{p}}^{\infty} \frac{dy}{\left(eBy+p_{z}^{2}\right)^{i+\frac{1}{2}}} \, .
\end{equation}
For the summation $T_S$, we similarly separate the low-$n$ terms from the infinite tail:
\begin{equation}
T_{S} = \sum_{n=1}^{m_{p}}F\left(p_{z}^{2}+neB\right) + \sum_{n=m_{p}+1}^{\infty} \sum_{i}\frac{c_{i}}{\left(neB+p_{z}^{2}\right)^{i+\frac{1}{2}}} \, .
\end{equation}
The infinite tail summation can be mapped analytically to the Hurwitz zeta function $\zeta(s, q) = \sum_{n=0}^{\infty} (n+q)^{-s}$. By completing the full sum from $n=1$ to $\infty$ and subtracting the lower finite segment, we find:
\begin{equation}
\sum_{n=m_{p}+1}^{\infty}\frac{1}{\left(neB+p_{z}^{2}\right)^{i+\frac{1}{2}}} = \frac{1}{\left(eB\right)^{i+\frac{1}{2}}}\zeta\left(i+\frac{1}{2},1+\frac{p_{z}^{2}}{eB}\right) - \sum_{n=1}^{m_{p}}\frac{1}{\left(neB+p_{z}^{2}\right)^{i+\frac{1}{2}}} \, .
\end{equation}
Replacing this back into $T_S$ yields a numerically regularized sum:
\begin{equation}
T_{S} = \sum_{n=1}^{m_{p}}\left[ F\left(p_{z}^{2}+neB\right) - \sum_{i}\frac{c_{i}}{\left(neB+p_{z}^{2}\right)^{i+\frac{1}{2}}} \right] + \sum_{i}\frac{c_{i}}{\left(eB\right)^{i+\frac{1}{2}}}\zeta\left(i+\frac{1}{2},1+\frac{p_{z}^{2}}{eB}\right) \, .
\end{equation}
Combining the finalized expressions for $T_S$ and $T_I$, the total potential $\Omega_{W}$ reduces to:
\begin{align}
\Omega_{W} = \frac{eB}{2\pi^{2}}\int_{-\infty}^{+\infty}dp_{z}\Bigg\{ 
&\sum_{n=1}^{m_{p}}\left[F\left(p_{z}^{2}+neB\right)-\sum_{i}\frac{c_{i}}{\left(neB+p_{z}^{2}\right)^{i+\frac{1}{2}}}\right] \nonumber \\
&+\sum_{i}\frac{c_{i}}{\left(eB\right)^{i+\frac{1}{2}}}\zeta\left(i+\frac{1}{2},1+\frac{p_{z}^{2}}{eB}\right) \nonumber \\
&-\left[\int_{0}^{m_{p}}dy~F\left(eBy+p_{z}^{2}\right)+\sum_{i} c_i \int_{m_{p}}^{\infty}\frac{dy}{\left(eBy+p_{z}^{2}\right)^{i+\frac{1}{2}}}\right] \Bigg\} \, .
\end{align}
\section[\appendixname~\thesection]{Explicit forms of gap equations} 
\subsection[\appendixname~\thesubsection]{Gap equation for $\Delta = 0$ and $eB \neq 0$}
In the presence of a diquark coupling, the thermodynamic potential is formally given by Equation~\eqref{OmegaT0}. Nevertheless, in the presence of an external magnetic field, the $\Delta \to 0$ limit requires special attention, since the normal-phase thermodynamic potential cannot be trivially recovered from the superconducting expression. Therefore, in this regime, the appropriate thermodynamic potential is given by~\cite{Ebert:1999ht,Ebert:2003yk}
\begin{eqnarray}
\Omega_{T=0} = \frac{(M-m_c)^2}{4G_S}
+ \frac{N_c}{4\pi^2} \sum_{f=u}^{d} \sum_{n=0}^{\infty}
\alpha_n |q_f| B
\int_{-\infty}^{+\infty} dp_z , E_{p,n}(M,B) \textcolor{black}{+ \Omega_{\rm med}},
\label{VeffNJLmu0B}
\end{eqnarray}
where the dispersion relation is defined as $E_{p,n}(M,B)=\sqrt{p_z^2 + 2n|q_f|B + M^2}$, \textcolor{black}{and 
\begin{eqnarray}
    \Omega_{\text{med}} = - \frac{N_c}{2\pi} \sum_{f=u}^{d} \sum_{n=0}^{\infty} \alpha_n |q_f| B
\int_{-\infty}^{+\infty} \frac{dp_z}{2\pi} [\mu_f - E_{p,n}(M,B) ] \theta[\mu_f - E_{p,n}(M,B)],    
\end{eqnarray}
is a finite $T\to0$ contribution.} In Equation~\eqref{VeffNJLmu0B}, the sum over $f$ runs over the quark flavors $u$ and $d$, with electric charges $q_u = 2/3$ and  $q_d = -1/3$, respectively.

After applying the MFIR procedure, Equation~\eqref{VeffNJLmu0B} can be rewritten as the sum of a \textcolor{black}{vacuum-regularized, a purely magnetic, and a medium contribution}:
\begin{eqnarray}
\Omega_{T=0} = \frac{(M-m_c)^2}{4G_S} + \Omega_{\text{Reg}} + \Omega_{\text{Mag}} \textcolor{black}{+ \Omega_{\text{med}}},
\end{eqnarray}
where
\begin{eqnarray}
\Omega_{\text{Reg}}  = - 2N_c N_f \int \frac{d^3p}{(2\pi)^3} E_p,
\label{OmegaReg}
\end{eqnarray}
with $E_p = \sqrt{\vec{p}^2 + M^2}$, and
\begin{eqnarray}
\Omega_{\text{Mag}} = -\sum_{f=u}^{d} \frac{N_c (|q_f|B)^2}{2\pi^2} \left[ \zeta'(-1,x_f) - \frac{1}{2}(x_f^2 - x_f)\ln (x_f) + \frac{x^2_f}{4} \right],
\label{OmegaMag}
\end{eqnarray} 
with $x_f = \frac{M^2}{2|q_f| B}$, analogously to the variable introduced in the $\Omega_1$ contribution discussed in Section~\ref{sec:model}.

The corresponding mass gap equation is then
\vspace{-6pt}
\centering 
\begin{eqnarray}
\frac{\partial \Omega_{T=0}}{\partial M} &=&
\frac{M - m_c}{2 G_S}
- 2N_c N_f M I_{M} \nonumber\\
&-& \sum_{f=u}^{d} \frac{M |q_f| B N_c }{\pi^2}
\left[
\ln \Gamma(x_f)
- \frac{1}{2}\ln(2\pi)
+ x_f
- \frac{1}{2}(x_f-1)\ln x_f
\right] 
\nonumber\\ 
&+& \textcolor{black}{ \frac{N_c}{2\pi} \sum_{f=u}^{d} \sum_{n=0}^{\infty} M\alpha_n |q_f| B
\int_{-\infty}^{+\infty} \frac{dp_z}{2\pi} \frac{1}{E_{p,n}(M,B)} \theta[\mu_f - E_{p,n}(M,B)]}
= 0,
\end{eqnarray}
where $I_M$ assumes different forms depending on the regularization scheme, and, for the MSS prescription, is given in Equation~\eqref{Imd0}.
\subsection[\appendixname~\thesubsection]{Gap equations for $\Delta \neq 0$ and $eB \neq 0$}

By employing Eq.~\eqref{OmegaT0} together with the contributions associated with the rotated charges within the MFIR framework,  presented in Eqs.~\eqref{W0}, \eqref{W1}, and \eqref{W05}, we explicitly derive the gap equations for magnetized, color-superconducting quark matter. The  constituent quark mass gap equation is given by
\begin{equation}
\frac{\partial \Omega_{T=0}}{\partial M} = \frac{M-m_c}{2 G_S} - \frac{\partial \Omega_0}{\partial M} - \frac{\partial \Omega_1}{\partial M} - \frac{\partial \Omega_{\frac{1}{2}}}{\partial M}=0 \, .
\end{equation}
 The definitions of $\chi$, $p_B$, $F(z)$, and $x$ are identical to those  introduced in Sec.~\ref{sec:MFIR},  and

\begin{align}
\frac{\partial \Omega_0}{\partial M}
&=
4 M I_{M,0}
+
\frac{M}{2\pi^2}
\left\{
p_f \sqrt{p_f^2 + M^2}
-\frac{M^2}{2}
\log\bigg[
\bigg(
\frac{p_f+\sqrt{p_f^2+M^2}}{M}
\bigg)^2
\bigg]
\right\}
\theta(\mu-M),
\\
\frac{\partial \Omega_1}{\partial M}
&=
-M\frac{eB}{2\pi^{2}}
\left\{
\chi-\frac{1}{2}\log(2\pi)
-\frac{1}{2}(2\chi-1)\log(\chi)
+\log[\Gamma(\chi)]
\right\}
\nonumber\\
&\quad
+\frac{eB}{2\pi^{2}}M
\sum_{n=0}^{p_{B,\max}}
\left\{
\alpha_n
\log\left[
\frac{\mu+\sqrt{\mu^{2}-p_{B}^{2}}}{p_B}
\right]
\theta\!\left(\mu-\sqrt{M^{2}+2eBn}\right)
\right\},
\\
\frac{\partial \Omega_{\frac12}}{\partial M}
&=
4MI_{M,\frac12}
+
\frac{eB}{4\pi^2}
\int_{-\infty}^{+\infty}
dp_z\,
\frac{\partial F(p_z^2)}{\partial M}
\nonumber\\
&\quad
-
M\frac{eB}{\pi^2}
\left\{
x-\frac12\log(2\pi)
-\frac12(2x-1)\log(x)
+\log[\Gamma(x)]
\right\}
\nonumber\\ 
&\quad + \frac{eB}{\pi^{2}}\int_{0}^{\infty} dp_{z} \Bigg\{ \begin{aligned}[t]
    & \sum_{n=1}^{m_{p}} \left[ \frac{\partial F\left(p_{z}^{2}+neB\right)}{\partial M} - \sum_{i} \frac{\partial c_{i}}{\partial M} \frac{1}{\left(neB+p_{z}^{2}\right)^{i+\frac{1}{2}}} \right] \\
    & + \sum_{i} \frac{\partial c_{i}}{\partial M} \frac{1}{\left(eB\right)^{i+\frac{1}{2}}} \zeta\left(i+\frac{1}{2}, 1+\frac{p_{z}^{2}}{eB}\right) \\
    & - \left[ \int_{0}^{m_{p}} dy \, \left[ \frac{\partial F\left(eBy+p_{z}^{2}\right)}{\partial M} \right] + \int_{m_{p}}^{\infty} dy \left( \sum_{i} \frac{\partial c_{i}}{\partial M} \frac{1}{\left(eBy+p_{z}^{2}\right)^{i+\frac{1}{2}}} \right) \right] \Bigg\} \, .
\end{aligned}
\end{align}
The explicit forms of $I_{M,0}$ and $I_{M,\frac{1}{2}}$ depend on the chosen regularization scheme;  for the MSS prescription, they are given in Eqs.~\eqref{Imd0} and~\eqref{Imd05}.

The diquark condensate gap equation is given by
\begin{equation}
\frac{\partial \Omega_{T=0}}{\partial \Delta} = \frac{\Delta}{2 G_D}  - \frac{\partial \Omega_{\frac{1}{2}}}{\partial \Delta} = 0 \, ,
\label{DeltaGapEq}
\end{equation}
since $\frac{\partial \Omega_0}{\partial \Delta} = \frac{\partial \Omega_1}{\partial \Delta}=0$. One obtains

\begin{align}
\frac{\partial \Omega_{\frac{1}{2}}}{\partial \Delta} 
&= 4 \Delta I_{\Delta} + \frac{eB}{4 \pi^2} \int_{-\infty}^{+\infty} dp_z \frac{\partial F(p^2_z)}{\partial \Delta} \nonumber \\
&\quad -\Delta\frac{eB}{\pi^{2}}\left\{ x-\frac{1}{2}\log(2\pi)-\frac{1}{2}(2x-1)\log(x)+\log[\Gamma(x)]\right\} \nonumber \\
&\quad + \frac{eB}{\pi^{2}}\int_{0}^{\infty}dp_{z}\Bigg\{ \begin{aligned}[t]
    & \sum_{n=1}^{m_{p}}\left[\frac{\partial F\left(p_{z}^{2}+neB\right)}{\partial\Delta}-\sum_{i}\frac{\partial c_{i}}{\partial\Delta}\frac{1}{\left(neB+p_{z}^{2}\right)^{i+\frac{1}{2}}}\right] \\
    & + \sum_{i}\frac{\partial c_{i}}{\partial\Delta}\frac{1}{\left(eB\right)^{i+\frac{1}{2}}}\zeta\left(i+\frac{1}{2},1+\frac{p_{z}^{2}}{eB}\right) \\
    & - \left[\int_{0}^{m_{p}}dy\,\left[\frac{\partial F\left(eBy+p_{z}^{2}\right)}{\partial\Delta}\right]+\int_{m_{p}}^{\infty}dy\left(\sum_{i}\frac{\partial c_{i}}{\partial\Delta}\frac{1}{\left(eBy+p_{z}^{2}\right)^{i+\frac{1}{2}}}\right)\right]\Bigg\} \, ,
\end{aligned}
\end{align}
where the  explicit form of $I_{\Delta}$ depends on the regularization procedure,  and, for the MSS scheme, it is given in Eq.~\eqref{idd05}.

\bibliography{ref}

\end{document}